\documentstyle[11pt,aasms4,epsf]{article}

\def\folio{\ifnum\pageno<2\nopagenumbers\else\number\pageno\fi}
\newtoks\headline \headline={\hss\twelverm\folio\hss} 
\newtoks\footline \footline={{\hfil}} 
\font\mathbf=cmmib10 scaled 1000             

\def\ref{\par\noindent\hangindent=2pc \hangafter=1 }
\def\amin{\ifmmode^{\prime}\else$^{\prime}$\fi}
\def\asec{\ifmmode^{\prime\prime}\else$^{\prime\prime}$\fi}

\def\cappage #1 #2 #3 {\vfill\eject\pageno=#1
\vglue 10 true in minus 10 true in \noindent{\bf Figure #2.} #3}
\def\ee #1 {\times 10^{#1}}
\def\ut #1 #2 { \, \hbox{#1}^{#2}}
\def\u #1 { \, \hbox{#1}}

\def\msol{\, \hbox{$\hbox{M}_\odot$}}
\def\kms {\, \hbox{km}\,\hbox{s}^{-1}}

\let\grad=\nabla
\def\cross{{\bf \times}}
\def\curl #1 {\grad \cross #1}
\def\div #1 {\grad \cdot #1}

\tighten
\doublespace

\def\msol   {\hbox{$M_\odot$}}                  
\def\kms    {\hbox{km{\hskip0.1em}s$^{-1}$}}    

\begin{document}

\title{A $\lambda$20cm Survey  of the Galactic Center Region I: 
Detection of Numerous Linear Filaments}

\author{F. Yusef-Zadeh}
\affil{Department of Physics and Astronomy, Northwestern University,
Evanston, Il. 60208 (zadeh@northwestern.edu)}

\author{J. Hewitt}
\affil{Department of Physics and Astronomy, Northwestern University,
Evanston, Il. 60208 (j-hewitt@northwestern.edu)}

\author{W. Cotton}
\affil{National Radio Astronomy Observatory, Charlottesville, VA 22903
(bcotton@nrao.edu)}

\begin{abstract}

This is a first in a series of papers presenting a sensitive $\lambda$20cm
VLA continuum survey of the Galactic center region using new and archival
data based on multi-configuration observations taken with relatively
uniform {\it uv} coverage.  The high dynamic range images cover the
regions within $-2^\circ < l < 5^\circ$ and $-40' < b < 40'$ with a
spatial resolution of $\approx30''$ and 10$''$.  The wide field imaging
technique is used to construct a low-resolution mosaic of 40 overlapping
pointings. The mosaic image includes the Effelsburg observations filling
the low spatial frequency {\it uv} data.  We also present
high resolution images of
twenty three overlapping fields using DnC and CnB array configurations.
These high-resolution images are sensitive to both compact and extended
continuum features with a wide range of angular scales with rms noise of
0.2 mJy beam$^{-1}$ in the outer parts of the Galactic center region.
The survey has
resulted in a catalog of 345  discrete sources as well as 140
images
revealing structural details of HII regions, SNRs, pulsar wind nebulae 
and more than 80 linear
filaments distributed toward the complex region of the Galactic center.

These observations show the evidence for an order of magnitude increase in
the number of faint linear filaments with typical lengths of few
arcminutes. Many of the filaments show morphological characteristics
similar to the Galactic center nonthermal radio filaments (NRFs). The
linear filaments are not isolated but are generally clustered in star
forming regions where prominent NRFs had been detected previously.  The
extensions of many of these linear filaments appear to terminate at either
a compact source or a resolved shell-like thermal source.  A relationship
between the filaments, the compact and extended thermal sources as well
as a lack of preferred orientation for many RFs  should
constrain models that are proposed to explain the origin of nonthermal
radio filaments in the Galactic center.

\end{abstract}

\keywords{ISM: Clouds---ISM: general---supernova
remnants-- HII regions: ISM --surveys}

\section{Introduction}

There is considerable interest in the physical processes occurring in
the Galactic center region where a number of unique thermal and
nonthermal radio continuum sources are hosted including the bright
nonthermal  compact radio source Sgr A$^*$. This compact source coincides
with a
concentration of dark matter considered to be a massive black hole at
the dynamical center of the Galaxy (e.g. Eckart \& Genzel 1996; Ghez et
al. 1998).  This interest has lead to a number of studies of star
formation activity in this region as well as  to a study of the origin 
of  unusual nonthermal activities. 
 Radio continuum structures have been studied with
single-dish and interferometric measurements on scale-sizes ranging
between a few degrees to sub-arcsecond resolutions over the last two
decades (Sofue and Handa 1984; Reich, Sofue \& F\"urst 1987; Handa et
al. 1987; Seiradakis et al. 1989; Haynes et al. 1992;  
Gray 1994a,b; LaRosa et al. 2000).

A number of recent large-scale interferometric studies of the Galactic
center have  been carried out with the Molongolo Observatory
Synthesis Telescope (MOST),  the Very Large Array (VLA) of the
National Radio Astronomy Observatory\footnote{The National Radio
Astronomy Observatory is a facility of the National Science Foundation,
operated under a cooperative agreement by Associated Universities, Inc.}
as well as the Giant Meterwave Radio Telescope (GMRT). MOST observations
at 843 MHz surveyed the region between $l = \pm5{^\circ}$ and $b =
\pm2^{\circ}$ and detected a number of supernova remnant (SNR)
candidates as well as ``the Snake'', a nonthermal radio filament (NRF)
near the Galactic center. Their analysis showed in excess of 14 SNR
candidates toward the Galactic center (Gray et al. 1991; Gray 1994a,b),
some of which have been confirmed at 610 MHz by GMRT observations (Roy
and Rao 2002; Bhatnagar 2002). In another study, large scale imaging of
the inner four  degrees of the Galactic center region was carried out
using one VLA pointing at 327 MHz (Anantharamaiah et al. 1991; Pedlar et
al. 1989; LaRosa et al. 2000; Nord et al. 2003) as well as at 74 MHz
(Brogan et al. 2003).  Nord et al. (2003) used wide field imaging
technique and detected additional linear filaments and determined
radio spectra of a number of radio sources at low frequencies. More
recently, polarization study of the linear filaments found originally 
at 327 MHz
showed that three of these  features 
are linearly polarized at  5GHz (LaRosa et al. 2004).
VLA observations at 1.4 GHz have
also been made with a number of antenna pointings but proper mosaicing
of these fields have not been carried out previously (e.g. Liszt 1985,
1992; Liszt and Spiker 1995).

Motivated by the ability to use wide field imaging technique to correct
non-coplanar effects, as was done at 327 MHz (LaRosa et al. 2000), we
present radio images of the Galactic center region at 1.4 GHz based on
archival and new observations taken in the last twenty years. 
We believe the data presented here can be combined
with other multi-wavelength observations to enhance our understanding of
this complex region of the Galaxy.  A more detailed analysis of the data
will be described elsewhere;  here,  we concentrate on the morphology
and
distribution of filamentary structures in the surveyed region.  A wide
variety of linear filaments with knotty but  continuous appearance are
reported. We argue that most of the new filaments are likely to be emitting
synchrotron radiation, are of the same class of NRFs and are associated
with star forming regions. In particular, we present evidence of thermal 
extended sources  and/or compact sources located at the extensions of the
terminus of these filaments.  Such a relationship should constrain models
that are proposed to explain the origin of nonthermal radio filaments in
the Galactic center.

The structure of the paper is as follows:  we first describe observations 
and data reductions followed by  presenting  
 a 20cm mosaic image with a resolution of 30$''$.
Because of a rich collection of radio sources in this region,  
the mosaic image is divided into eight segments, each of which 
is displayed  as contours and grayscale images. 
In section 4, we discuss 
high-resolution ($\approx10''$) images of Sgr C (G359.5-0.0) and its
extensions, Sgr E (G358.7-0.0), 
the  radio Arc (G0.2-0.0) and its extensions, Sgr A (G0.0-0.0), Sgr
B1
(G0.5-0.0), Sgr B2 (G0.07-0.0) and radio continuum sources 
distributed between Sgr B2 and l$\approx5^0$.  
A catalog of point sources  using the high resolution data
has also been created.  Earlier catalog of  small
diameter sources over a
much larger region of the inner Galaxy has been presented based on
snapshot observations with the VLA at 20cm by 
Zoonematkermani et al. (1990). Since the emission at 20cm is
dominated by extended features in the Galactic plane of the Galactic
center region, our survey data sample the {\it uv}   
plane reasonably well and therefore 
is more
complete and more sensitive to  identify the point sources distributed
within the inner two degrees of the Galactic center region than 
earlier studies. 

\section{Observations and Data Reductions}
\subsection{Low-resolution data}

We have used 40 overlapping $\lambda$20cm fields based on archival and
new VLA observations in its compact, hybrid array configuration.  Nine
of the pointings were carried out on January 30, 2003 using the compact
DnC array configuration of the VLA.
Figure 1a shows a schematic diagram of all
the pointings represented by black and gray circles of diameter 30$'$
corresponding to the 20cm primary beam. The individual pointings have
different {\it uv} coverage but the combined data sample the {\it uv}
plane reasonably well. In fact, none of the selected archival data were
based on snapshot observations shorter than 30 minutes.  Field 34 has
been observed using the DnC array but in the HI spectral line mode with
excellent {\it uv} coverage;  the continuum channel was selected for
this survey data. 
Each individual data set was
edited, calibrated and self-calibrated before the final images were
constructed.  

Prior to the construction of the final low-resolution
mosaic image, the individual images were convolved to 30$''$ and were
incorporated with single-dish observations based on Effelsburg data
(Reich 1982; Reich and Reich 1986) in
order to account for the lack of low spatial frequency {\it uv} data. 
Single dish images and those derived from different VLA
configurations were combined using the "feathering" technique.
The feathering technique consists of making a weighted average of
the images to be combined in the spatial frequency (Fourier transform)
plane.  
The weighting is designed to produce a combined image which
incorporates the spatial frequencies from the most relevant 
image at each location in the spatial frequency plane and with a
smooth transition between images. 
The relevant range of spatial frequencies for a given image is
determined by the configuration of the array used to make an
interferometric image or the size of the single dish and can be
parameterized by the beam size.
The images being combined should collectively and adequately sample all
spatial frequencies out to that corresponding to the full resolution
of the combined image.
After combination in the Fourier domain, the image is then
transformed back to the image domain.
In the implementation used here, the weights in the Fourier domain
have a constant value except for a Gaussian profile hole in the center
where the 
Gaussian is the Fourier transform of the (Gaussian) beam of
the next lowest resolution image.
Since the images being combined are restored CLEAN or single dish
images, they have already been tapered at spatial frequencies not
sampled by the instrument making the image.
Thus, the limited resolution of an instrument should not be reflected
in the spatial frequency weighting function used for its image.
The combined image is normalized by repeating the process replacing the
input images by images of a unit point source convolved to the same
resolution. 
This technique very effectively selects
the portions of the {\it uv}-plane best sampled in each of the input
images.
A proper mosaic image of the overlapping $\lambda$20
cm fields  was
constructed using FLATN in AIPS.

Table 1 gives a listing of the antenna pointings in celestial and
galactic coordinates and the rms noise of each individual image which
has been combined with the Effelsburg data. The rms noise of the
low-resolution images corresponding to the fields
outside the central degree of the Galactic center
is $\approx$0.5 mJy beam$^{-1}$
when the single-dish data has not been added; the rms noise  increases
by a
factor of
20 in the central 0.5 degrees toward the Sgr A region.

\subsection{High-resolution data}

Additionally, we have constructed a mosaic image with the highest resolution
available for a given pointing. Most of the individual sources have
resolutions about 10$''$.  The high resolution data are carried out in the
hybrid DnC and CnB array configurations but there are also few pointed
observations in the D, C, B and A arrays.  The single-dish data has 
not been added to these images with the exception  of the region between 
l$\approx1^0$ and 5$^0$ where the VLA and Bonn images are feathered
together, as seen in Figure 26.
Table 2 gives a listing of 23
antenna pointings in celestial and galactic coordinates and the array
configurations that were employed.  
Column 1 in both Tables 1a,b corresponds to  the field number in the
decreasing 
order of Galactic longitude. 
Figure 1b displays a schematic diagram of
all the high-resolution pointings, similar to that shown in Figure 1a.  
Multi-configuration data are combined in the {\it uv} plane, self-calibrated
before the final images are constructed.  Sgr A, Sgr B and the
Radio Arc (G0.2-0.0) 
have
high-resolution A-array data and ``feathering'' technique was used 
in  combining the data in the image plane. In order to bring
out small-scale faint features, some of the final images were
constructed by
restricting the low-spatial frequency data to $\it {uv} >
0.4 k\lambda$.    This cut-off of the inner {\it uv} plane 
suppresses the strong background
emission arising from the Galactic center region.

The criterion  that we selected  to identify the linear filaments
was   by  detecting
them either in more than one 20m field of view or 
in images taken in two different array configurations of the VLA. 
In addition, the length-to-width ratio of the filaments had to be  
greater than few in order to avoid shell-like structures such as  HII
regions
with sharp edges. 
All the observed fields
overlapped
with each other and were observed in more than one array configuration.  
In addition, because of the spatial coherence of the long filaments, 
we were able to identify them   easily even though   
the surface brightness of some of the faint filaments was similar to
the background rms level. The probability of positive
detection of these faint filaments increases as their lengths extend 
over  many pixel sizes.

Table 3 lists all the filaments described in this paper. Column 1 shows
the sources within which the filaments are found or associated with,  in
the order in which they are
described here.  Columns 2 and 3 show the nomenclature used   on
figures and the corresponding Galactic coordinates. Column 4
shows whether the filaments are identified as NRFs or 
 considered as candidate NRFs. Using 
high resolution images, 
the length, the orientation and 
the integrated flux density of background-subtracted  filaments 
are given in columns 5, 6
and 7,
 respectively. 
The flux densities of
the filaments that are very faint and/or suffer from a
negative
bowl caused by the lack of low-spatial frequency {\it uv}  data have not
been estimated. Column 8 show  the references to 
the  previously identified  radio filaments. 

The compact point sources detected in high-resolution data are fitted
 with a two-dimensional single Gaussian using SAD in
AIPS. Columns 1 to
8 of Table 4 lists the galactic and celestial coordinates, the peak and
integrated flux, the size and the corresponding errors, respectively,
for each compact source found in the high resolution data.  Column 9
shows the field numbers (as shown in Figure 1b)  corresponding to the
high resolution data where compact sources are fitted with
signal-to-noise greater than 5. column 10 indicates a flag given by
SAD when there is a high point in the residual image of the fit, thus
the fit is poor. The
sources that are unresolved in this table are blanked in columns 6 -- 8.  
A total of  345	  compact sources were identified as listed in the
point
source catalog, all of which have sizes less than 1$'$. A catalog of
extended compact sources beyond 1$'$ scale length as well as the
analysis of the  point source catalog  will be
given elsewhere.

\section{Low-resolution $\lambda$20cm Mosaic Image}

Figure 2a shows a mosaic image of the central region between $-5^{\circ}
< l < 2^{\circ}$ and $-40' < b < 40'$ along the Galactic plane.  The
high dynamic range image with a resolution of 30$''$ shows the bright
saturated central region (in black color)  where Sgr A, B, C and the Arc
are located; the inclusion of single-dish data has accounted for the
low spatial frequency visibility data. The images that do not include
the low spatial frequency {\it uv} data produce artifacts around 
bright and extended continuum sources. Figure 2b shows a close-up view
of the central Sgr A
region. The low-resolution mosaic image is displayed in eight different
grayscale and contour panels in Figures 3 to 10. With the addition of
low-spatial frequency {\it uv} data, 
the background continuum emission varies between 20
and 40 mJy beam$^{-1}$  across the surveyed region.

\section{High-resolution $\lambda$20cm Continuum Images}

In the following sections, the large-scale distribution of
high-resolution images are first displayed before close-up views of
individual sources are shown in grayscales and contours.  A brief
description of the new filamentary structures is described.  Unlike the
low-resolution data, the single-dish data have not been folded into the
high-resolution VLA data.  We first concentrate on the bright sources on
the negative galactic longitudes before we show the images of the Arc,
Sgr A and Sgr B. Finally, we present a high resolution mosaic image of
radio features distributed at high positive longitudes beyond Sgr B2
(G0.7-0.0).

\subsection{Sgr C (G359.5-0.0) \& its Northern and  Southern Extensions}
\subsubsection{Large-scale View}

Sgr C  lies at the western "footprint"  of a large-scale
radio structure (aka ``The Galactic center lobe'' or G359.4-0.5) that
has been detected
using single-dish observations of the Galactic center (e.g., Sofue and
Handa 1984). Figure 9a shows a low-resolution image of Sgr C and its
northern extension known as the western Galactic center lobe.  This
diffuse
and extended structure runs away from the galactic longitudes ranging
between 
l$\approx-0.5^0$ and --0.2$^0$ toward positive galactic latitudes.  
The large-scale diffuse western Galactic center lobe appears to
coincide with AFGL 5376 (Uchida, Morris and Serabyn 1990). 

Sgr C
hosts one of the most prominent
radio continuum sources in the Galactic center region with its nonthermal
and thermal radio continuum components (Liszt 1992; Liszt and Spiker
1995). Multiple filaments of Sgr C appear to end abruptly inside a
molecular cloud HII complex with a velocity of --65 \kms (Liszt 1992;
Liszt and Spiker 1995). Figure 11a shows the high-resolution image of Sgr
C and  a large number of labeled linear features in
its vicinity.  
Figure 11b, which is based only on
B-array observations, shows a large, southern  view of the Sgr C region.
The fields to the northwest and southwest of Sgr C are shown in 
Figures 11c and 11d where prominent NRFs known as G359.54+0.18 (the ripple
filament)  and G359.1-0.2 (the Snake filament) are displayed. The
southern extension of the
Snake  is shown in Figure 11e.  This figure  shows 
the field containing the southern end of the Snake,  SNR
G359.1-0.5, SNR
G359.0-0.9 (LaRosa et al. 2000; Bamba et al. 2000)  and 
G359.23-0.82 (the Mouse) within which a young pulsar has recently been
found  
(Yusef-Zadeh and Bally 1987; Camilo et al. 2002). Highlights of some 
of the features found in these fields are described below.

\subsubsection{Individual Images}

A large number of  new NRF candidates are detected throughout
Sgr C and its
northern and southern extensions.  
Table 3 lists a total of 33 filaments associated with the Sgr C region. 
All these filaments are distributed
within the large-scale western Galactic center lobe which appears to
have a counterpart at 
 8$\mu$m and 20$\mu$m (Bland-Hawthorn \& Cohen  2003; 
Uchida, Morris and Serabyn 1990).
The relationship
between the filaments of Sgr C and the Galactic center lobe, AFGL 5356 is
not well  understood. However, the
presence of these linear features may suggest that thermal and
nonthermal features in this region  may have the same
origin. A number of the new filaments are faint with surface
brightness 1--5 mJy beam $^{-1}$ and do not
appear to be
preferentially oriented perpendicular to the Galactic plane. 
The names of the filaments, as indicated  in Table 3 with their Galactic
coordinates, are labeled on Figures 11 and 12. 

\begin{description}
\item[G359.45-0.06 (RF-C1)]
The most
prominent nonthermal radio filaments (RF-C1 in Table 3) run along the
eastern edge of the Sgr C thermal source.  
Figures
12a,b show close-up views of the Sgr C HII region and the NRFs (RF-C1 in
Fig. 11a) as they cross the HII region. We note that the vertical
nonthermal filaments of Sgr C become brighter and narrower as they curve 
toward the diffuse
semi-circular HII
region at
its base near G359.57--0.067.  
Also a point source at a
level of $\approx$ 5 mJy beam$^{-1}$ is located at one end of a
subfilament in RF-C1 near G359.43+0.03. 
Similar behavior has been
detected in the NRFs of the radio Arc as
the NRFs  cross the HII regions associated with the Arched filaments
(Yusef-Zadeh and Morris 1988).  These images show morphological evidence
for the likely association of the nonthermal radio filaments of Sgr C and
its HII region. The contours of Sgr C filament in Figure 12b also
indicates that while the filaments are enveloped by weak continuous
emission, the core of the filamentary structure
appears  bright and  knotty as 
 the filaments extent toward the Sgr C HII region (G359.43-0.09).

\item[G359.54+0.18 (RF-C3)] Another
well-known nonthermal system of filaments is G359.54+0.18 (aka "the ripple
filament") which lies to the north of Sgr C.  The ripple filament resolves
into multiple parallel components with a terminus that flares in the
direction toward the Galactic plane (Bally and Yusef-Zadeh 1989;
Yusef-Zadeh, Wardle and Parastaran 1997).  Figures 11c and 12n show
grayscale images of this nonthermal source which extends for
about 15$'$ (36 pc at the distance of 8.5 kpc). This sharp drop-off in
the brightness distribution of the
filament in the direction away from the Galactic plane is reminiscent of
the network of filaments in  the Arc as they become diffuse and
faint in the direction normal to the Galactic
plane. Earlier observations had detected the
brightest component of G359.54+0.18 which extends for about 5$'$. The
newly detected diffuse emission extending to the northwest of
G359.54+0.18 has a typical surface brightness of $\sim$ 1 mJy
beam$^{-1}$ 
which is   weaker 
than the the brightest segment of RF-C3 by a
factor of 20. Evidence for such a large-scale nonthermal filament
indicates that 
the western Galactic center lobe has a nonthermal component.

\item[G359.49-0.12 (RF-C4):]
 Another system of filaments (RF-C4 in Figures 11a and 12d) appears to
act as a bridge between two diffuse features at G359.47-0.17 and
G359.57-0.07. This system breaks up into at least three components
aligned along its long axis. A faint diffuse emission at a level of 0.2
mJy beam$^{-1}$, though not displayed in Figure 12d, appears to connect
the southwestern component to G359.47-0.17.  If these components are
part of a coherent filamentary structure, the brightness of the
filament must vary by a factor of 5 to 10 along its 6$'$ extent, as seen
in Figure 12d. We believe the knotty structure of G359.49-0.12 consists
of two filaments running parallel to each other but with
non-uniform surface brightness along their lengths.  The lack of faint diffuse
emission between the northeastern and southwestern filaments may be due
to the missing low spatial frequency {\it uv} data which has not been
added to high resolution images.  Although no diffuse emission has been
detected between the individual filaments, we believe the alignment of
these linear features suggests that they have the same origin.  Similar
alignment of NRFs has been detected in the Arc as the filaments cross
G0.18-0.04 (Yusef-Zadeh and Morris 1987b).

\item[G359.44+0.01 (RF-C5)] Figure 12c shows a close-up view of the
northern extension of Sgr C where a curved filamentary structure (RF-C5)
appears to cross the Sgr C filaments (RF-C1).

\item[G359.34-0.15 (RF-C13)] Figures 11a and 12e show two different
views of RF-C13 which consists of multiple parallel filaments
brightening at G359.32-0.16 where they are curved most. A diffuse
feature G359.27-0.22 is noted in Figure 12e at the southwestern
extension of this system of filamentary structure. We also note another
filamentary feature G359.32-0.18 (RF-C29), as seen Figure 12e, 
  with an
extent of
$\approx2'$ and a position
angle of 34$^0$ to the south of
RF-C13
at G359.32-0.18.

\item[G359.21+0.54 (RF-C16)] This new filament, as shown in Figure 11c,
extends for about 10$'$ and is one of the narrowest (width of
$\approx10''$) and most uniform filaments found in the Galactic center
region in terms of its surface brightness ($\approx1$ mJy beam$^{-1}$).  
The southern extension of this filament, when extrapolated, appears to
cross a point source at G359.26+0.42 (see Table 4). The filament is
aligned along the direction of the western component of the Galactic
center lobe.

\item[G359.44+0.44+0.14 (RF-C6)  and G359.37+0.11 (RF-C11) ] 
There are
two sets of pair of
filaments RF-C6 \& RF-C7 as well as RF-C11 \& C12 which appear to cross
each other. The pair of RF-C6 and C7, as shown in Figure 11c and 12p,
run perpendicular and parallel to the Galactic plane whereas RF-C11 and
RF-C12, as labeled in Figure 11c, cross each other at a 60$^0$ angle.
RF-C7 in Figure 12p appears to show wiggles along its horizontal
direction. 
The extension of RF-C12, as shown  in Figure 11c,  appears to
terminate at the compact source G359.24+0.16.  The full extent of RF-C12
extends for about 10$'$ but the brightest portion has a length of
$\sim2'$ with a surface brightness of 1 mJy beam$^{-1}$. Although the
brightness of faint emission between RF-C12 and G359.24+0.16 is at a
level of background noise (0.1 mJy beam$^{-1}$), we  believe  that
the faint portion of RF-C12 is a coherent structure and is an  extension
of the
brightest portion of the filament. 

\item[G359.37-0.03 (RF-C10)] Similar to
G359.45-0.06 (RF-C1) and other
filaments near Sgr C such as 
G359.59-0.17 (RF-C2), G359.34-0.15 (RF-C13) and G359.29+0.11 (RF-C15), 
G359.37-0.03 (RF-C10)  appears to bend and join a diffuse
source G359.36+0.0, as shown in Figure 12i.  The faint diffuse emission at
the terminus of
G359.37-0.03 (RF-C10) resembles the HII region associated with
G359.45-0.06 (NRF-C1).  

\item[G359.32-0.06 (RF-C14)]  Figure 11a and 12q show 
the appearance of a ``bipolar''  feature with a length of $\approx45''$
as it extends to the northeast
and southwest. 
G359.32-0.06 (RF-C14) is identified as one of many 
linear
features with
a position angle ranging roughly between 30$^0$  and  60$^0$, as seen in
Figure 11a. 
These filaments are grouped together running  parallel to each other
in the Galactic plane with similar orientations to
the  individual components of G359.49-0.12 (RF-C4). 
However, the filaments grouped together near G359.32-0.06 (RF-C14)
appear to be
broader and surrounded by diffuse extended features along the Galactic
plane. 

\item[Additional  Filaments] Figures 12f--s show contour
and grayscale images of a
subset of linear filaments found  in the vicinity of the Sgr C region. 
A  variety of linear filaments 
designated as G359.37+0.11 (RF-C10),  G359.41-0.26
(RF-C17),
G359.35-0.24 (RF-C18), G358.98-0.25 (RF-C22), 
G359.48-0.23 (RF-C24), G359.70-0.28 (RF-C26),  G359.30-0.06 (RF-C32) and
G359.63+0.04 (RF-C33) are displayed in these figures (see
Table 3).
The brightness distribution of these filaments
generally show knot-like structure with non-uniform surface brightness as 
seen in G359.1-0.2 or the Snake filament,  as described below in more
detail. 
Some of the filaments appear to terminate  
in the vicinity of  a compact
source such as RF-C24 in Figure 12h.


\item[G359.23-0.82 (the Mouse) and G359.28-0.26] Figure 13a shows
contours of the
nonthermal pulsar wind nebula G359.23-0.82 (Yusef-Zadeh and Bally
1992; Camilo et
al. 2002) whereas Figure 13b displays a bright and extended cometary
thermal source G359.28-0.26 (Uchida et al. 1992); This HII source has 
a bright
counterpart in the \emph{Midcourse Space Experiment} (\emph{MSX})
8$\mu$m image.

\item[G359.47+0.03 (RF-C28)] Figure 13c shows a
number of HII regions  in
Sgr C. A compact HII feature G359.65-0.09 appears at
the terminus of G359.47+0.03 (RF-C28), as seen  
 Figure 13c.

\item[G359.1-0.2 (NRF-C19, the Snake)] The region to the southwest of
Sgr
C lies a prominent nonthermal filamentary structure known as the Snake
filament extending for more than 20$'$.  Figure 11d shows this striking
feature with its two prominent
kinks near G359.13-0.20 and G359.12-0.26 (e.g., Gray et al. 1991, 1995).
The northern extension of G359.16-0.04 coincides with a thermal source
G359.16-0.04 (see also Figures 9a,b) whereas the southern end of the
Snake appears to cross SNR G359.1-0.5 (Caswell and Haynes 1987; Uchida
et al. 1996).  The prominent HII region G359.16-0.04 is 
known to have a
number of compact radio continuum
sources (Uchida et al. 1996).  
Figures 12l,m show close-up views of
two kinks noted along the length of the Snake. There are two compact
sources G359.13-0.20 and G359.12-0.26 which are located within or in the
vicinity of the region where the surface brightness of the Snake
increases.  The Snake is known to show a gradient in its spectral index
at the location of the kinks (Gray et al. 1995).  The spectral index
along the NRFs is generally constant and flat between 6 and 20cm with
the exception of the  kink at G359.13-0.2 where the spectrum steepens to
$\alpha$=--0.5. Subfilamentation has also been detected in the vicinity
of the kinks (Gray et al. 1995). 
Similar to the Sgr C
NRF (RF-C1 in Table 3), the Snake has  a continuous structure but shows
knotty
appearance along its length. The brightest segments of the Snake are in
the middle where the filaments are most curved. Like a number of 
long isolated linear filaments, the Snake is likely to be made up 
of  a pair of 
filaments that run parallel to each other or is a hollow 
limb-brightened cylindrical structure. A a more detailed study 
of this structure 
will be 
given elsewhere.

\end{description}

\subsection{Sgr E (G358.7-0.0)}

\subsubsection{Large-scale View}

High resolution radio continuum and radio recombination line observations
of Sgr E, one of the least studied HII complexes in the Galactic center
region, show that the distribution of the continuum emission is dominated
by compact sources displaying thermal characteristics (Liszt 1992; Gray et
al. 1993; Cram et al. 1996).  Figures 14a,b show high and low-resolution
20cm images of Sgr E, respectively.  These images show a number of bright
compact sources amounting to 2$\times10^3$ $\msol$ of ionized gas excited
by about 20 OB stars (Gray et al. 1993). There are two linear filaments
G358.62+0.07 (RF-E1) and G3558.79-0.13 (RF-E2) listed in Table 3.  
These filaments, which appear to have similar morphology to NRFs, are
displayed in Figure 14a.  There are other filamentary structures seen in
this image and  they are difficult to separate from  thermal features
associated with
extended ionized shells in this HII complex.

The low-resolution image of Figure 14b shows a string of HII regions
distributed on a partial ring;  this distribution is reminiscent of HII
regions seen in Sgr B2 and W49N (e.g., De Pree et al. 2000).  We note a
large-scale diffuse filamentary structure G358.60-0.27 (RF-E3) arising
from the center of Sgr C. The closest source lying at one end of the
filament is Sgr E53 at G358.3-0.12 (Gray et al. 1993). The nature of this
new
elongated feature (RF-E3 in Table 3) extending for about 20$'$ is
unknown.
Its width is about 1$'$ and its length is as long as the length of the
Snake but its structure is highly distorted along its length.
Similar to the Snake and Sgr C filaments, one end of G358.60-0.27 
appears
to terminate near the center of the Sgr E HII complex.

\subsubsection{Individual Images}

\begin{description} 
\item[G358.62+0.07 (RF-E1)]
 Figures 15a shows contour
images of a filamentary structure RF-E1 (the ``spoon'' filament) with a
position angle of 117$^0$. This filament  appears to be associated 
with a thermal source
G358.63+0.06 which is designated as Sgr E19 by Cram et al. (1996).  The
width of this filament
increases away from the direction of Sgr E19 along its $\sim100''$ extent.
The morphology of the filament indicates that the filament is originated 
with the HII region.

Radio recombination line emission from this bright and compact source
confirms its thermal nature but at a radial velocity near 0 \kms which is
considerably different than the velocity of molecular and ionized gas
associated with Sgr E (Cram et al. 1996).  This is perhaps one of the best
examples that a thermal source lies at the terminus of a filamentary
structure.  Similar to RF-C1 and its Sgr C HII region, high resolution
image of the spoon filament runs tangent to an asymmetric shell-like
thermal source with an opening that is not aligned along the orientation
of RF-E1.  The integrated flux density of the compact source and the
filament at 20cm are 409 and 36 mJy, respectively (see Table 3).

\item[G358.79-0.13 (RF-E2)]
Figure 15b shows
grayscale contours of the central region of Sgr E. 
Another filamentary feature RF-E2 or
G358.79-0.13 is detected
at the center of a diffuse circular feature. The  diameter of
this circular feature is  3.4$'$
having 
surface brightness of $\approx$ 1 mJy beam$^{-1}$.
 The morphology of
RF-E2 is similar to RF-E1 in that the filament appears to join the bright
thermal  feature G358.72-0.12 (Cram et al. 1996). The position angle of
RF-E2
is close to 97$^0$. We note that a number of  diffuse features in Sgr
E  are  elongated in the direction  approximately parallel to
those of  RF-E1 and E2.
\end{description}

\subsection{The Radio Arc (G0.2-0.0) and its Extensions}

G0.2-0.0 known as the radio continuum Arc is the
prototype filamentary structure
consisting of a network of vertical filaments with lengths of about 30 pc
distributed symmetrically with respect to the Galactic equator
(Yusef-Zadeh, Morris and Chance 1984; Yusef-Zadeh and Morris 1987a,b,c;
Anantharamaiah et al. 1991).  
 This region contains the
largest concentration of long and bright NRFs as well as three
prominent HII regions:  
G0.07+0.04 (the Arched filaments), G0.18-0.04 (the Sickle)  and
G0.16-0.06
(the Pistol). 
The nonthermal  bundle of filaments,  called RF-S0 in Table 3,  
cross  all three HII regions, as  labeled on Figure 16a. 
These HII regions are thought to be ionized by two massive
clusters of hot stars known as the Arches and the Quintuplet clusters
(e.g. Cotera et al.  1996; Serabyn, Shupe and Figer 1998; Figer et al.
2002).   Additional  curved structures,   similar to the 
Arched filaments,  though much fainter,  are also detected on  the
negative latitude side of the 
linear filaments of the radio Arc. These curved structures 
 appear to cross the  nonthermal filaments near G0.16-0.15.  


Like Sgr C and its extensions, the Arc and its extensions lie at the
eastern ``footprint" of the  Galactic center lobe (Tsuboi et al. 1986).  
Two polarized lobes of emission are detected along the extensions of the
vertical filaments of the Arc away from the Galactic plane  (Tsuboi et al.
1986). The vertical network
of filaments in the Arc extend toward positive and negative latitudes
--0.75$^0<b<+0.75^0$ as the filaments become diffuse, and faint in
their surface brightness (Sofue and Handa 1984). 
Here, we show 20cm images of the southern and northern extensions of the
Arc.

\subsubsection{The Southern   Extension of G0.2-0.0}

\subsubsection{Large-scale View}

Figures 16b,c show the grayscale continuum images of two
overlapping $\lambda$20cm
fields located along the southern extension of the Arc. The new filaments
are labeled  on this figure and their Galactic coordinates are listed
under S. Arc  in
Table 3. A number of filaments to the south and southeast of the Arc 
has been recognized to have nonthermal characteristics based on 
single-dish polarization measurements (Tsuboi et al. 1986; Seiradakis 
et al. 1985) and 6cm VLA observations (Yusef-Zadeh 1986).  
The long filament G0.17-0.42 (RF-S5)  becomes
brightest
in the middle and breaks up into two parallel filaments as they run toward
negative latitudes. This filament appears to be distinct from the network
of filaments associated with the Arc.  The filament G0.11-0.40 (RF-S7)
which is more diffuse and fainter than
RF-S5 runs also in the same direction. A low-resolution version of RF-S5
and
S7 in Figure 7a shows that these filaments coincide with the
eastern and western edges of an
 elongated  structure with a width of about 3$'$ and a length of about
20$'$ running perpendicular to the Galactic plane. There is clearly 
diffuse linearly polarized emission arising from the region between
RF-S5 and RF-S7 as detected in 
the southern component  of  the Galactic center lobe (Seiradakis et 
al.1985; Tsuboi et 
al. 1986). The orientations of
RF-S5 and S7 are tilted by about 20$^{\circ}$ with respect to the position 
angle
of the network of filaments that comprise the Arc, as seen in Figure
16a.

\subsubsection{Individual Images}

\begin{description}

\item[G0.17-0.42 (RF-S5)]
Figures 17a,b show grayscale and contour images of RF-S5. 
We note a bright
compact source G0.19-0.69 along the southern extension of RF-S5 in Figure
17b whereas a bow shock-like structure G0.13-0.28 is noted at the
northern
extension of RF-S5 in Figure 17a.

\item[G0.13-0.58 (RF-S6)]
A new  linear feature RF-S6 with its wiggly structure  appears
to cross
RF-S5 in Figure 16b. The
southwestern and northeastern extensions of this weak filamentary
structure, as shown in Figure 17c, appear to be terminated at  an
unresolved compact source G0.08-0.61 at one end and an extended bright
feature G0.28-0.48 at the other end, respectively. Because of the 
complex  extended emission noted in this region, the weak distorted
filament
RF-S6 
crossing RF-S5 needs to
be confirmed in future observations.  
Figures 17d,e show
two renditions of G0.28-0.48 with its core and a partial shell   structure. 
A linear structure with an orientation parallel to the Galactic plane  
appears to be terminated at the  core of G0.28-0.48.

\item[G0.39-0.12 (RF-S2)]
Another filamentary structure whose northern and southern extensions
terminate at a diffuse feature 
and  a compact source is G0.39-0.12 (RF-S2). Figure 17f shows the
compact source
G0.42-0.32 and the diffuse feature  G0.38+0.02; both these sources 
are aligned along the extensions of RF-S2. Similarly, G0.43-0.31
(RF-S1) consists of two parallel filaments whose northwestern extensions  
appear to terminate 
at two compact sources G0.32-0.19 and G0.32-0.2. A contour map  
of these compact sources is shown in Figure 17m. 

\item[G359.88-0.28 (RF-S9), G0.30-0.27 (RF-S3)]
Figures 17g-l show
contour and grayscale representations  of additional  filaments found in
the region  south of the radio Arc.  In particular, we note that
G359.88-0.28 (RF-S9), 
as seen in Figures 17j and k, is a
filamentary structure that appears to have a similar morphology to that
of Sgr C where one end of the  filament terminates at  a thermal source  
G359.0--0.32. A
higher resolution image  of RF-S9 is displayed
in Figure 17k. Another bundle of filaments G0.30-0.27 (RF-S3) can be
seen in Figures 16b and
17i where two pairs of filaments running parallel to each other. We also
note additional isolated filament RF-S8 with 
an extent of 3.7$'$ in
Figure 17l.  

\item[G0.09-0.09 (RF-S10)]

Figures 17n shows a relatively broad filament G0.09-0.09 running roughly
perpendicular to the Galactic plane. This filament which was recently
identified as G0.087-0.087 at 32 GHz is linearly polarized (Reich 2003)
and is thought to lie at the western edge of a dense molecular cloud
G0.13-0.13. The eastern and western edges  of this molecular cloud show
X-ray filamentary structure; these  edges are thought to
be
signifying interaction 
sites of the molecular cloud G0.13-0.13 with the filaments of the Arc. 
The eastern side at molecular cloud at G0.13-0.12 and with 
G0.09-0.09 (RF-S10) lies on the 
western side of the cloud whereas the filaments at 
G359.0.08-0.09 lie at the eastern side of the cloud (Tsuboi,
Ukita and Handa 1997; Yusef-Zadeh, Law and Wardle 2002).  Figure 16a
shows a dearth of continuum emission between RF-S10 and the network of
the filaments of the Arc where the CS molecular cloud is distributed.

\item[G0.06-0.08 (RF-S11), G0.09-0.15 (RF-S12)]

Two other linear filaments in the vicinity of RF-S10 are shown
in Figures 17n through 17p. G0.06-0.08 appears to have the same 
orientation as RF-S10 but displaced about two arcminutes west of RF-S10. 
G0.09-0.15 is a  faint filament with typical  surface
brightness 
of 1 mJy beam$^{-1}$ and appears to cross the 
southern filaments of the Arc,  as seen in  Figure 16a.   We also
note an elongated  compact source to the south of RF-S12 which
is oriented 
in the same direction as that of  RF-S12. 

\item[G0.2-0.0 (NRFs-S0)]

The nonthermal radio filaments of the Arc at G0.2-0.0 cross the Galactic
plane where two prominent HII regions G0.18-0.04 (The Sickle nebula) and
G0.16-0.06 (the Pistol nebula) are distributed. Figures 17q,r show
high-resolution grayscale and
contour images of the region where the filaments of RF-S0 cross these
HII sources. Morphological arguments have been made
previously that the HII regions and
the nonthermal filaments are associated with each other (Yusef-Zadeh and
Morris 1987b).

\end{description}

Figures 18a-f show grayscale contours of non-filamentary 
extended
features distributed to the south of the radio Arc at G0.2-0.0; many of
the individual 
features presented in these figures are also labeled  in  Figure
16b.

\subsubsection{The Northern    Extension of G0.2-0.0}
\subsubsection{Large-scale View}

Similar to its southern extension, the northern extension of
the radio Arc near l$\sim0.2^0$ lies within a  large-scale
diffuse  structure  running perpendicular to the Galactic
plane (e.g.
Tsuboi et al. 1986; Seiradakis et al. 1985).  The extension of the NRFs
of the Arc toward positive latitudes has been argued to be physically
associated with a lobe of polarized emission (Yusef-Zadeh and Morris
1987c). This lobe is also identified as the eastern ``leg'' or
``footprint''  of the Galactic
center lobe crossing the Galactic plane at l$\sim0.2^0$ (Sofue and Handa
1984). Figure 19a shows
a low-resolution image of the northern extension of the Arc as a group 
of NRFs,   designated as RF-N1,  appear to pass through G0.17+0.15
(Yusef-Zadeh and Morris 1988).  We note G0.15+0.23 (RF-N1) and 
G0.08+0.15 (RF-N2)  cross
the thermal
Arched filaments of the Arc G0.07+0.04.
The Arched
filaments  H II complex is comprised of a series of curved, narrow
ridges
of radio emission which define the northern  edge of the well-known
Galactic center radio Arc (Yusef-Zadeh, Morris, \& Chance 1984;
Morris and Yusef-Zadeh 1989; Lang, Morris and Goss 2002). The
Arched filaments extend for 9$'\times6'$ (or 23 $\times$× 16 pc at the 
Galactic center distance of 8.5 kpc) and are located
10$'$ in projection from Sgr A$^*$.

Figure 19b-d show  high and low-resolution images of a number of filaments
found  
at positive latitudes between l$\sim-0.5^0$ and $0.2^0$. 
The
brightest segment
of Sgr A, as shown in white in Figures 19b,c,  is  known as Sgr A East.
A
collection  
of  broad linear  features, aka the ``streamers'',  appears to arise from
Sgr A as it bends at 
positive galactic latitudes b$\sim3'$. We note  faint and  narrow linear
 filaments G359.93+0.07 (RF-N6) running  along  the direction of the
streamers. 
These  new linear features as well as the streamers terminate 
within the ionized gas of Sgr A West which itself is surrounded  
by the circumnuclear molecular ring. The Sgr A West HII region is 
known to be powered by the UV radiation from the IRS 16 young star
cluster.

\subsubsection{Individual Images}

\begin {description}  
\item[G0.15+0.23 (RF-N1), G0.08+0.15 (RF-N2)]

Figures 20a,b show   contour and grayscale images  of RF-N1 and
RF-N2. The  system of 
linear 
filaments in RF-N1 is most distorted as it crosses the
shell-like
structure G0.17+0.15 (Yusef-Zadeh and Morris 1988). 
The system of RF-N1 appears to be comprised of three distinct filaments, 
two of which surround an HII region G0.17+0.15. 
 Like many filaments
described above (e.g. RF-C1
and C4 of Sgr C in Figure 11a), these RFs also run along the edge of the
shell-like thermal source G0.17+0.15. 
We believe the morphology of the filaments indicate strongly that 
these NRFs are interacting with G0.17+0.15 which is known to have an 
8$\mu$m {\it MSX} counterpart.  
  
Figure 20b shows a lobe of nonthermal diffuse emission detected to the
north of
N1 and
N2. Polarization measurements of the lobe show that the  magnetic field
is 
oriented perpendicular to the Galactic plane similar to the 
orientation of the NRFs N1 and N2 (Tsuboi et al. 1986).
Figure 20c shows the surface brightness of RF-N2 peaks in the middle as it
curves toward higher latitudes before the filament becomes broader and
more diffuse. The width of RF-N2 changes by about a factor of 2 as its
surface brightness decreases by a factor of $\approx5$.  This filament
appears to terminate at the location where a compact source G0.12+0.32 is
located,  as seen in Figure 20a. 

\item[G0.02+0.04 (RF-N3), G359.98+0.02 (RF-N4)]
Figures 20d,e show detailed structures of
two shell-like thermal sources  G0.02+0.04 and G--0.012+0.12, aka
H4 and H1,
respectively (Yusef-Zadeh and Morris 1987; Zhao et al. 1993; Lang et al.
1999). 
These HII regions appears to lie at the terminus of two RFs N3 and
N4. The incomplete shell-type HII region G0.02+0.04 has an opening along
the direction of RF-N3 whereas RF-N4 appears to run along the bright edge
of G--0.012+0.12.  Similar to G0.117+0.15, the morphological
association of HII region/filament in G0.02+0.04 is compelling. Radio
recombination line H110$\alpha$ emission has been
detected from G0.02+0.04 showing a radial velocity of --39 \kms with low
electron temperature of 3.5$\times10^3$K. 

\item[G359.96+0.09 (RF-N5), G359.93+0.07 (RF-N6)] 

 A larger view of RFs N4, N5 and
N6 can be viewed in Figure 20f where  G359.96+0.09 (RF-N5)
appears to cross the
streamers of
Sgr A West as well as G359.98+0.02 (RF-N4). G35996+0.09 consists of
two parallel filaments, one of which continues to extend toward
RF-N4.   It is not clear if RF-N4 and RF-N5 are associated with each
other. 

\item[G359.79+0.17 (RF-N8), G359.88+0.20  (RF-N7)]

Figures 20g-h show close-up  grayscale images of G359.79+0.17 (RF-N8)
and 
G359.88+0.20 (RF-N7).  G359.79+0.17  comprises of multiple
parallel filaments whose intensity peak at midpoint along
its curved structure. The filaments appear to be separated
from each other as they run toward southeast. The extension of the
fainter filament
to the southwest appears to terminate at a  compact source G359.80+0.06. 
  
The elongated structure of the bright source 
G359.87+0.18 in Figure 20h is an artifact of the  distortion caused 
possibly  by
phase and
amplitude errors. 
This source G359.87+0.18 is identified as an extragalactic FR II radio
galaxy (Lazio et
al. 1999). In spite of the  errors affecting the appearance of 
G359.87+0.18, RF-N7 is clearly identified
as an isolated linear radio filament adjacent to this source. 

\item[G0.09-0.05 (RF-N9), G0.12-0.01 (RF-N13)]

Figure 20i shows a close-up view of the Arched filaments with a
resolution of 2.1$''\times1.2''$. We note a linear filament
RF-N9 with a position angle of 171$^0$ originating
from a diffuse feature G0.08+0.01, resembling a number of thermal 
ionized features associated with the Arched filaments. 
The faint
linear filament 
with a surface brightness of 1-2 mJy beam$^{-1}$ extends for about
3$'$.  Figure 20j shows contours of G0.08+0.01. 
 Another weak linear structure G0.12-0.01 running almost parallel to the
Galactic
plane is detected in Figure 20i. The surface brightness of this feature
G0.12-0.01 (RF-N13) is about 1 mJy beam$^{-1}$ as it extends for about 
1$'$ with a position angle 90$^0$.

The extension of G0.09-0.05 to the south, as  best seen in Figures 
20k and 19c,  terminates at a pair of compact sources at G0.08+0.09
and G0.08+0.09
having  
peak flux densities  of 9.45 and 6.4 mJy beam$^{-1}$, respectively. 
These high-resolution images clearly show that G0.08+0.15 (RF-N2) and
G0.09-0.05 (RF-N9) are
distinct from each other but both filaments have very similar
orientations, 
as seen in Figures 20i,k.
Although  the extension of RF-N2 to the
north 
is very prominent (see Figure 19a),  its southern extension 
is faint in its surface brightness and appears to 
terminate at a bright compact source at G0.01-0.05 with a peak flux
density of 14.9 mJy beam$^{-1}$. We also note  enhanced emission at 
the position of G0.08+0.02 along 
the extension  of RF-N2, as seen  in Figure 20i. RF-N2  crosses 
thermal Arched filaments, as seen  in  Figure 20j. There are other weak
linear 
filaments that follow the orientation of the Arched filaments but are
difficult to separate them from the bulk of thermal emission arising 
from  the Arched filaments. 
\end{description}

\subsection{Sgr A (G0.0+0.0) and its Neighbors}
\subsubsection{Large-scale View}

Multi-wavelength observations of the Galactic center region shows a clumpy
molecular ring with a scale of 2 to 5 pcs rotating around Sgr A$^*$. The
ring is heated by a centrally concentrated cluster of hot, young stars, 
the IRS 16 cluster, 
with similar characteristics to the Arches and the Quintuplet clusters.  
Within the ring's central cavity, three ``arms'' of ionized gas (Sgr A
West) are generally in orbital motion around the center.  On a larger
scale, a non-thermal structure (Sgr A East or SNR G0.0+0.0) is projected
against Sgr A
West and the molecular ring.
Sgr A East is considered  to be a shell-type supernova remnant. 
The Sgr A complex is comprised  of both thermal Sgr
A West and nonthermal  Sgr A East as well as a halo 
of nonthermal gas  enveloping both Sgr A East and West
(e.g., Yusef-Zadeh and Morris 1987a; Pedlar et al. 1989).  

Two other radio continuum sources that have thermal spectrum are 
G359.7-0.0 and G359.8-0.30, both of which are  located in projection
within the region 
between Sgr A and Sgr C. A number of compact thermal sources have  been
detected in G359.7-0.0 (Liszt and Spiker 1995). Detailed radio continuum 
study  of G359.8-0.30 has  not been carried out previously.
Low-resolution images,  as seen 
in Figures 8a,b, show a large shell-like structure with a diameter of 
$\sim10'$ centered roughly around G359.8-0.30. An 8$\mu$m {\it MSX} image 
of this region  suggests that G359.8-0.30  has thermal characteristics.

\subsubsection{Individual Images}

\begin {description}  

\item[G359.98-0.03 (RF-A7), G359.96-0.03 (RF-A10)]

Figure 21a shows a close-up view of the shell-type SNR G0.0+0.0 (Sgr A
East) which is  elongated
along the Galactic plane. The three-armed spiral structure of Sgr A
West G359.94-0.05 as well as a cluster of
HII regions G359.98-0.07 distributed at the southeastern edge of Sgr A
East are also noted. 
 A number of straight and curved filamentary
structures with a range angular sizes  between
30$''$ to 1$'$ is detected throughout Sgr A East; The new filaments  are
labeled on
Figure 21a. In particular, we note a number of filamentary features
designated as G359.98-0.03  (RF-A7), G359.96-0.03 (RF-A10), 
G359.96-0.04 (RF-A11), 
G359.95-0.05
(RF-A12) and G359.96-0.05 (RF-A13).  These filamentary structures 
(RFs A7, A10, A12 and A13) give 
the appearance of an 
elliptical shell with a diameter 2$'\times1'$ centered on G359.96-0.04.  

Because of the wide range of angular scales of radio features found in
the complex  region surrounding Sgr A, we present a number of figures
which are sensitive to bringing out features with  scale
lengths ranging between few arcseconds to few arcminutes, as  displayed
in Figures 21b,c,d. The large-scale ``streamers''
are noted to run toward more positive Galactic latitudes, as seen in
Figures
21c,d.  The base of the streamers terminate where
the Sgr A West HII
region lies.  A different rendition of the Sgr A streamers designated as
G359.93+0.07 (RF-N6) is  also shown in Figures 19b,c and 20f.  Like Sgr
C
and the radio
Arc that show prominent vertical filaments away from the Galactic plane,
Sgr A is also recognized to have a long, straight nonthermal filament
G359.96+0.09 (RF-N5 in Figure 21d).   G359.96+0.09 (aka "thread") 
appears to terminate near the
eastern edge of the streamers (Morris and Yusef-Zadeh 1985; Yusef-Zadeh
and Morris 1987). G359.96+0.09 (RF-N5) extends for about ten arcminutes
toward positive
galactic latitudes. This filament is resolved into two subfilaments and
becomes very faint compared to the bright shell-like HII regions
distributed between the
Arc and the Sgr A complex.  G359.96+0.09 is brightest at its  midpoint
 where the linear filamentary structure   appears to deviate from a
straight line as 
it  breaks up into
two components.
 
\item[G359.88-0.07 (RF-A4), G359.85-0.02 (RF-A6)]

Figure 21d
shows a number of filaments with different orientations and curvature.  
G359.88-0.07 (RF-A4) and G359.85-0.02 (RF-A6)  run 
perpendicular and parallel to the Galactic 
plane. We also note  G359.89-0.10 (RF-A3) consists of two components
that 
deviate from a straight geometry. 

\item[G359.78+0.01 (RF-A16), G35973+0.02 (RF-A17)]

In the region to the west of Sgr A where G359.7-0.0 HII complex lies,
we find two faint filaments G359.78+0.01 (RF-A16) and  G359.73+0.02
(RF-A17) running almost perpendicular to the Galactic plane. 
Figures 21d,e show  these faint filaments. We note that 
the extensions of RF-A16, RF-A17 and  G359.72+0.03 (RF-A18)  terminate at
three resolved
features labeled as G359.79-0.04, G359.74-0.02 and G359.72-0.01,
respectively.

\item[G359.72-0.50 (RF-A15), G359.68-0.39 (RF-A19)]
The region to the southwest of Sgr A lies a large-scale diffuse structure
G359.8-0.30
showing  a number of diffuse and compact sources. Figures 21f through 21h
show two  
different representations of G359.8-0.30 where we find a number of 
linear structures. In particular, a long linear structure G359.72-0.50
(RF-15) extending for about 15$'$ with a  position angle of 59$^0$, as
listed
in Table 3. G359.68-0.39 (RF-A19) is another long filament that runs 
perpendicular to the Galactic plane. Two diffuse  blob-like structures 
G359.7-0.4 and G359.6-0.2 are surrounded by a number of linear structures  
RF-C2, RF-C26. RF-A19 and RF-A15. 

\end {description}

Figures 22a-h display contour distribution of some of the filamentary
structures found to the west of Sgr A, Sgr A East and G359.8-0.30. 
 The filaments are oriented with
a wide variety of position angles with respect to the Galactic plane.
Figure 22c show two linear filaments oriented parallel (RF-A6) and almost
perpendicular (RF-A5) to the Galactic plane, respectively. 
Figures
22b,e show distorted filamentary structures (RFs A2, A3 and A8).  
Distorted filaments are generally characterized to have multiple
components and are generally brighter than  the surface brightness of
single isolated  filaments. For
example, curved filaments  RFs-A2, A3 in Figure 21b  consist of
multiple parallel components and appear to be brighter in their
surface brightness than single filaments such RFs-A5, A6 and A9.  
Some filaments also show bow shock-like
structures like the system of filaments identified as RF-A7 in
Figures 21a
and 22f.  Lastly, we note that some of the filaments lie adjacent to
compact sources. In particular, RF-A4 lies within 45$''$ of two
compact sources, G359.87-0.04 and G359.87-0.09 (Figure 22b).  The latter
source is
known as a thermal  source  based on its flat spectrum (Ho et al.
1985) and  is resolved into two components,  as seen in Figure 22b.

Figures 23a-b show contour images of Sgr West and 
the well-known  string of HII regions lying at the edge of Sgr A East.
Another  string of compact sources beyond the edge of Sgr A East is also
presented in Figure 23c. The southern arm of the spiral-shaped ionized gas 
associated with Sgr A West (Figure 23a) near G356.17-0.05 
shows a drop  in its  brightness distribution which is likely to be due
to  optical depth effect.  
Sgr A West is known to
be optically thick at 327 MHz but this is the first evidence that 
similar effect has  been detected at 20cm (a more detailed analysis 
will be given elsewhere). 

\subsection{Sgr B2 (G0.7-0.0) and its  Neighbors}

\subsubsection{Large-scale View}

Radio continuum observations of the well-known star forming region Sgr B
show two major components Sgr B1 (G0.5-0.0), Sgr B2 (G0.7-0.0) and an
intervening source G0.6-0.0 (e.g., Mehringer et al. 1992).  Figure 24a
shows a high resolution image revealing the large-scale distribution of
all three components.  Sharp semi-circular and  bar-like ionization
fronts are  noted in Sgr B1 whereas numerous ultracompact HII regions
with cometary morphology 
dominate the emission in Sgr B2 and G0.6-0.0.  Figure 24b, c show the same
region as Figure 24a except that only the highest and lowest resolution
data are used to construct the images, respectively. Two long
isolated filaments
G0.43+0.01 (RF-B1) and  G0.39+0.05 (RF-B2)   are revealed in
the region to the west of Sgr B, as shown in Figures 24c,d.  We also note
a number of extended  HII
features north of G0.6-0.0  at b$>2'$.

\subsubsection{Individual Images}
\begin {description}
\item[G0.43+0.01 (RF-B1), G0.39+0.05 (RF-B2)]

Two isolated filamentary structures, G0.43+0.01 (RF-B1) and G0.39+0.01
(RF-B2), are shown in Figure
24d. Unlike the bright filaments which deviate from straight line and show
multiple filamentary structures with curved morphology, these thin,
isolated, single
filaments are faint with uniform surface brightness of about 1 mJy
beam$^{-1}$. We also note a semi-circular-shaped thermal feature
G0.38-0.02 and two compact sources G0.41-0.02 and G0.42-0.06 distributed 
in the
vicinity of the filaments. Neither of these compact sources are aligned
with the extensions of the filaments to the south. However, it is possible
that these images are not sensitive to  the diffuse and faint segments of
the
filaments with surface brightness $<$ 1 mJy beam$^{-1}$ extending  to the
compact  sources. 
There are examples of radio filaments such as RF-N8 in
Figure 20g in which the filaments appear to bend close to a compact
source. A third linear filament that has been identified previously
in adjacent images is G0.39-0.12 (RF-S2)  which extend for about 8$'$ in
the direction
perpendicular to the Galactic plane. This filament  is broader than RF-B1
and RF-B2 but shows
the same orientation as the rest of filaments in Sgr B. The northern end
of G0.39-0.12  appears to
bend toward Sgr B1 before it dissapears in the bright continuum emission
from Sgr B1.

\end{description}

There are a number of linear features detected throughout the Sgr B region
whose continuum emission is dominated by a mixture of bright, compact,
thermal sources and diffuse and extended features.  We believe
that
these features do not belong to the population of nonthermal radio
filaments and they are likely to be thermal features associated with
ionization fronts at the interface of the HII and molecular clouds. Figure
25a shows a contour image of the Sgr B region.  Because of the
large-extent and the complex structural details of Sgr B, 
grayscale images of several segments of the image are displayed  
followed by their  contour representations.  High-resolution image are
divided into three sections;
the grayscale and contour images of each of these segment are presented in
Figures 25b-g.

\subsection{Continuum Features at  l$>0.7^0$}

\subsubsection{ Supernova Remnants}

Figure 26 shows a high-resolution mosaic image of the region between
l$\sim1^{\circ}$ and  $5^{\circ}$.  The continuum emission in this region
is
dominated by HII regions and SNRs. 
Figures 27a-d show  four  known SNRs G0.9+0.1, G1.4-0.1, 
G3.7-0.2 and G1.9+0.3 with new structural details (see Gray
1994a,b and the references therein).

\begin {description}
\item[SNR G0.9-0.1]

 SNR G0.9-0.1
appears to show a
symmetrical linear feature, as seen in Figure 27a, 
arising from its center where the bright pulsar wind nebula lies
(Helfand and Becker 1987). The northeastern half of this
barrel-shaped SNR which is facing away from the Galactic plane 
is brightest. 
The HII region G0.83+0.19 to the north of G0.9+0.1 has a
measured radial velocity of 9.7 \kms (Kuchar \& Clark 1997). 

\item[SNR G1.4-0.1]

SNR G1.4-0.1 in Figure 27b shows a symmetrical  shell structure with a 
number of compact sources projected against the remnant (see Table 4).
The southern half of this remnant is thought to be interacting 
with a molecular cloud inferred from the detection of OH (1720 MHz)
maser emission (Yusef-Zadeh et al. 1999; Bhatnagar 2002).   

\item[SNR G3.7-0.2]

Figure 27c shows the northern half of another  barrel-shaped SNR G3.7-0.2
whereas Figure 27d displays the crescent-shaped structure of SNR
G1.9+0.3. The   brightest half of this shell-type  remnant runs normal
to the
Galactic plane.   

\item[SNR G1.0-0.1]
  
Figure 27e   shows a low-resolution image of SNR G1.0-0.1. This 
remnant which is also called Sgr D remnant has a barrel-shaped
morphology with its northern  half facing the Galactic plane;  
the eastern half near G1.16-0.23  shows diffuse extended structure
which may be part of the remnant. 

\item[Candidate SNR G1.0-0.1]

A candidate SNR G3.7-0.1 is resolved into four components, as shown in  
Figure 27f; 
two of the brightest components are 
 G3.67-0.1
and
G3.66-0.1. Bhatnagar (2000) has recently modeled   the continuum
spectrum as a combination of thermal and nonthermal emission. If 
the spectrum contains nonthermal emission, the
 shell-like morphology of G3.66-0.1 is
consistent with being a supernova remnant.  
A hydrogen-recombination line velocity of
3 \kms based on low spatial resolution observations  has been measured
toward
G3.66-0.12 (Caswell \& Haynes 1987).

\end {description}
 
\subsubsection{Diffuse Thermal Sources}

High-resolution radio continuum images do not show any evidence of linear
filaments beyond Sgr B with the exception of one possible source G1.3+0.1.
This source, as shown in Figure 28a, consists of two shell-like features
that appear to be in contact with each other.  A linear feature with an
extent of about 4$'$ and width of 45$''$ appears to arise from the
location where the two incomplete shells meet.  The linear feature appears
broader than typical radio filaments, as catalogued in Table 3.  The
largest concentration of NRFs and RFs found in this survey are detected
mainly within the region confined by the Galactic center lobes. Other
examples of shell-like or cometary HII features are presented in Figures
28b-c.  Figures 28d,e show two extended HII complexes G4.57-0.12 and
G4.42+0.12 (Lockman et al. 1989). The thermal source G4.57-0.12 in Figure
28d shows an extended triple component 6$'$ in extent. A
hydrogen-recombination line velocity of +18 \kms has been measured
(Caswell and Haynes 1987). G4.27+0.04 in Figure 28f is resolved into a
shell and a compact source with a diameter of $\sim2'$ (Bhatnagar 2002).  
 Radial velocity measurement of HII region G2.51-0.03, as seen in
Figure 28m, indicates a velocity of 8.3 \kms (Lockman, et al. 1996).
A number of other extended, thermal features are noted between G0.7-0.0
and G4.5+0.0 as their grayscale representations are displayed in Figures
28g-n.


\section{Discussion}

The present 20cm survey has provided a  great deal of information on the
structural details  of HII regions, compact
sources,  nonthermal filaments  and supernova remnants in the complex
region of the 
Galactic center. Future multi-wavelength study of these features  
should further our understanding of these  objects. In addition,
the present 
survey    
 reports a catalog of more than 80 system of filamentary structures,
many of which remain as candidate  nonthermal filaments.
All the new filaments are distributed in the region between Sgr B1
(G0.5-0.0) and Sgr E (G358.7-0.0).  It is likely that additional system of
filaments could be uncovered with observations having higher sensitivity, 
a more uniform {\it uv} coverage and a larger spatial coverage  
at higher  galactic latitudes.  A
wide range of morphological
details are detected among the linear filaments.  Some appear as a single,
faint, straight and isolated filament with a surface brightness of 1 mJy
beam$^{-1}$ and others appear as bright network of two or more curved
filaments running parallel to each other. The largest concentration of
filaments are characterized in pairs separated  between  few
arcseconds to 30$''$ from each
other as they run parallel to each other. It is not clear if the pair of
closely spaced narrow filaments signify  two distinct parallel filaments
or a single limb-brightened filament.  We also note that some filaments
break
up into multiple subfilaments when they are most distorted (e.g., kinks).
Some filaments show a  curvature along their
lengths
as their brightness peak in midpoints as they gently curve.

Most of the new linear filaments reported here are likely to be of the
population of radio filaments known to have nonthermal characteristics as
recent 327 MHz and 6cm polarization measurements indicate. 
Some of the 20cm
linear filaments reported here are also found  independently
at 327 MHz (LaRosa et al. 2004; Nord et
al.  2004). Future
polarization and spectral index measurements should confirm the nonthermal
characteristics of the new filaments listed in Table 3.

Unlike earlier studies indicating bright filaments are continuous and
generally running perpendicular to the Galactic plane, we note a large
fraction of the filaments showing knot-like structure with gaps in between
and the presence of a wide range of orientations with respect to the
Galactic plane. We note several linear filaments parallel to the Galactic
plane similar to G358.85+0.47 (Lang et al. 1999) which was discovered as
the first filament not being perpendicular to the Galactic plane.  
Figure 29 shows a schematic diagram of the distribution of 
the filaments found in this survey. It is clear that the longest  
filaments roughly perpendicular to the Galactic plane.  
Unlike
the most prominent long filaments with an extent $>5'$, the  
short
filaments do not show a preferred orientation perpendicular
to the Galactic plane. 
The wide range of orientations of the linear filaments provide
strong observational constraints on the hypothesis that there is a
large-scale poloidal magnetic field threaded in the Galactic center
region. Assuming that the short filaments are nonthermal, these
observation are not inconsistent with  models that 
argue the origin of the magnetic field is local and dynamic (e.g., LaRosa,
Lazio and Kassim 2001; Yusef-Zadeh 2003).

We also find that a number of 
filaments  appear to terminate at an extended thermal source.  In
some cases, the linear filaments appear to run tangential to the edge of
thermal sources or arise directly from a shell-like HII region (e.g., 
G0.02+0.04 (RF-N3) in
Figure 20d). The physical relationship between diffuse thermal sources
and
the filaments need to be tested with additional observations. However, if
such a relationship exists  based
on morphological grounds, then we argue that the origin of the filaments
is in the prominent Galactic center  star forming regions.
It is not a coincident that the largest
concentration of the filaments are populated in star forming regions in
the Galactic center region.  Recent interpretation of the origin of the
filaments considers a mechanism in which the collective winds of massive
WR and OB stars within a dense stellar environment produce nonthermal
particles (Rosner and Bodo 1996; Yusef-Zadeh 2003). In the context of this
model, HII regions which are associated with radio filaments are ionized
by the young cluster of stars. Furthermore, we speculate that the low
electron temperature
measured toward some HII regions in the Galactic
center region might  be
explained by the contribution of nonthermal continuum emission
 which
results a low estimate of the line-to-continuum ratio. 
Also,  unusual kinematics  of some of the HII regions 
may be explained by the  dynamical interaction of 
supersonic motion of nonthermal filaments
with  HII regions.

We also note that compact radio sources are found along the extension of
the terminus of a large number of  filaments. If these compact sources 
are associated with the filaments, they may act as a hot spot 
in the context of a jet model where the supersonic motion of a jet-like 
filament is impacting the ISM.   
(Yusef-Zadeh and K\"onigl 2004). The present survey
shows that linear filaments have morphological
characteristics similar to
extragalactic FR I jets. 
Future high
resolution  and proper motion  study of these compact sources should be 
able to test the jet model. 
 
Lastly, since the ``footprints'' of the eastern and western lobes lie on
two prominent sources with a large number of filaments, as seen in Sgr C
and the Arc, the role and the relationship of the large-scale lobes should
be tied to the origin of  the nonthermal
filaments. It is striking to note that very few filamentary structures 
are found beyond the edges of the footprints of the Galactic center
lobe. It is possible the winds from clustered star formation in a
star burst environment are responsible for the origin of the filaments as
well as the lobes (Bland-Hawthorn and Cohen 2003; Law et al. 2004).

Acknowledgments: We thank  the referee for useful comments. 
This work was  partially funded by  the NRAO student support
program and NSF AST-03074234 and NASA NAG-9205.

\begin{figure}
\figcaption{ (a) A diagram showing the number of VLA antenna pointings
numbered in the order of decreasing Galactic longitude are  
used in this low-resolution (30$''$) $\lambda$20cm survey. Fields A, B,
C, D and E correspond to a number of overlapping pointings. (b) is
similar to (a) except that the antenna pointings are based on
high-resolution (10$''$) observations. Overlapping pointings are
represented by fields A, B and C  and are shown in light gray.}
\end{figure}

\begin{figure}
\figcaption{ (a) A mosaic image of the entire surveyed region at 
$\lambda$20cm with a resolution of 30$''$.
The logarithmic grayscale range is between  30.0 100.0 mJy beam$^{-1}$. 
(b) A close-up view of the 
brightest region toward Sgr A. Prominent nonthermal radio
filaments (NRFs), SNRs and HII regions are designated}
\end{figure}

\begin{figure}
\figcaption{ (a) A first segment of the mosaic image is shown with a
spatial resolution of FWHM=30$''\times30''$.  (b) The corresponding
contours are drawn with levels set at 35, 40, 45, 55, 70, 100, 150 and
250 mJy beam$^{-1}$. The dotted line near 3$^0$ 20$'$, 00 15$'$ is an
artifact of mosaicing. }
\end{figure}
\begin{figure}
\end{figure}

\begin{figure}
\figcaption{(a) A second  segment of the mosaic image is shown with a
spatial resolution of FWHM=30$''\times30''$.  (b) The corresponding
contours are drawn with levels set at 34, 37, 40, 43, 46, 50, 56, 64, 80
mJy beam$^{-1}$.
}
\end{figure}
\begin{figure}
\end{figure}

\begin{figure}
\figcaption{ (a) A third  segment of the mosaic image is shown with a
spatial resolution of FWHM=30$''\times30''$. There was a small overlap
between pointing number 9 and A, as shown in Figure 1a, thus the mosaic
image has cut off the region with a high noise level due to the primary
beam correction.  
(b) The corresponding
contours are drawn with levels set at 46, 52, 58, 64, 70, 76, 82, 90,
120, 155, 230 and
600 mJy beam$^{-1}$.
}
\end{figure}
\begin{figure}
\end{figure}

\begin{figure}
\figcaption{ (a) A fourth  segment of the mosaic image is shown with a
spatial resolution of FWHM=30$''\times30''$.  (b) The corresponding
contours are drawn with levels set at 40, 45, 50, 60, 70, 80, 90, 100,
130, 160, 200, 250, 300, 400, 600, 1000, 2000, 4000 and 
 8000 mJy beam$^{-1}$.}
\end{figure}
\begin{figure}
\end{figure}

\begin{figure}
\figcaption{ (a) A fifth  segment of the mosaic image is shown with a
spatial resolution of FWHM=30$''\times30''$.  (b) The corresponding
contours are drawn with levels set at 25, 30, 35, 40, 45, 50, 60, 70,
80, 90, 110, 130, 160, 200, 240, 320 and  500 mJy beam$^{-1}$.}
\end{figure}
\begin{figure}
\end{figure}

\begin{figure}
\figcaption{ (a) A sixth  segment of the mosaic image is shown with a
spatial resolution of FWHM=30$''\times30''$.  (b) The corresponding
contours are drawn with levels set at 20, 25, 30, 35, 40, 45, 50, 60,
70, 80, 90, 100, 120, 140, 160 and 180 mJy beam$^{-1}$.
}
\end{figure}
\begin{figure}
\end{figure}

\begin{figure}
\figcaption{(a) A seventh  segment of the mosaic image is shown with a
spatial resolution of FWHM=30$''\times30''$.  (b) The corresponding
contours are drawn with levels set at 45, 50, 55, 60, 65, 70, 75,
80, 90, 100, 110, 120, 140, 160, 180, 220, 300, 400 and  500
mJy beam$^{-1}$.
}
\end{figure}

\begin{figure}
\figcaption{(a) An eigth segment of the mosaic image is shown with a
spatial resolution of FWHM=30$''\times30''$.  (b) The corresponding
contours are drawn with levels set at 40, 42, 46, 48, 50, 54, 58, 62,
66, 70, 75, 80, 90, 110 and 160
mJy beam$^{-1}$.
}
\end{figure}
\clearpage
\begin{figure}
\figcaption{ (a) A grayscale distribution of Sgr C with a resolution of
8.1$''\times3.3''$ (PA=-11$^0$). The coordinates
of the labeled radio filaments are listed in Table 3.
RFs in this region are  labeled.
(b) The greyscale image of the southern extent of Sgr C where several
newly identified RFs are found. This image is based only on the
B-array configuration data with a resolution of
8.5$''\times3.5''$ (PA=-12$^0$).
(c) The grayscale image of the region to the north of Sgr C where the
prominent
G359.54+0.18 (RF-C3) lies, also known  as the ripple filament.
The resolution of this image is 10.1$''\times8.3''$ (PA=64$^0$).
(d) The region to the southwest of Sgr C lies the Snake or G359.1-0.2
with
a resolution of 12.1$''\times6.8''$ (PA=52$^0$). (e) The
southern extension of the Snake filament G359.1-0.2  as well as two
shell-type SNRs
 G359.1-0.5,  G359.0-0.9 and the Mouse G359.23-0.82
are shown with a resolution of
12.8$''\times8.4''$ (PA=56$^0$). }
\end{figure}

\begin{figure}
\figcaption{ (a-b) Grayscale and contour images,
respectively,  of the brightest
region of Sgr C with contours set at -0.25, .25, .5, 1, 2, 4, 6, 8,
 10, 12, 14, 16, 18, 20 mJy beam$^{-1}$ with a resolution of
8.1$''\times3.3''$ (PA=-11$^0$) and a greyscale flux range between
-4 and 20  mJy beam $^{-1}$; 
(c) Greyscale contours of RF-C1, C5, C8, C10, C11, and C12 at 1, 1.5,
2, 3, 4, 5, 10, 15, 20, 25, 30, 40, 50, 60, 70, 80, 90, and 100 mJy
 beam$^{-1}$ with a resolution of 8.1$''\times3.3''$ (PA=-11$^0$)
and a greyscale flux range between
-1 and 10  mJy beam $^{-1}$; (d)
Grayscale contours of RF-C4 with contours similar to (c); (e) Contours
of
RF-C13 with contours similar to (c) but
and a greyscale flux range between
-0.5 and 2  mJy beam $^{-1}$ ; (f) Contours of RF-C18,
C17 with
levels 0.25, 0.5, 1, 1.5, 2, 3, 4, 5, 10, 15, 20, 25, 30, 40, 50, 60,
70,
80, 90, and 100 mJy beam$^{-1}$ and 
and a greyscale flux range between
-0.4 and 4  mJy beam $^{-1}$;  (g) Contours of RF-C2 with levels
-0.5,
0.5, 0.75, 1, 1.25, 1.5, 2, 2.5, and 3 mJy beam$^{-1}$
and a greyscale flux range between
-1.9 and 5  mJy beam $^{-1}$;  (h) Contour
levels .75, 1, 1.25, 1.5, 2, 3, 4, 5, 10, 15, 20, 25, 30, 40, 50, 60,
70,
80, 90, and 100 mJy beam$^{-1}$ and a greyscale flux range between
-0.1 and 2 mJy beam $^{-1}$;  (i)
Contours of RF-C10 with levels
.75,
1, 1.25, 1.5, 2, 3, 4, 5, 10, 15, 20, 25, 30, 40, 50, 60, 70, 80, 90,
and
100 mJy beam$^{-1}$ and a greyscale flux range between
-0.1 and 4  mJy beam $^{-1}$;  (j) Contour
levels 0.5, 1, 2, 3, 4, 5, 6, 7, 8,
10,
15, 20, 25, 30, 35, 40, and 50 mJy beam$^{-1}$ and a greyscale flux
range between -1 and 4  mJy beam $^{-1}$; (k) Contours of RF-C26  
with levels -1, 1, 1.5, 2, 2.5, 3, 4, 5, and 6 mJy beam$^{-1}$
and a greyscale flux range between
-3 and 15  mJy beam $^{-1}$; (l) The
southern kink of the Snake with a resolution of 112.1$''\times6.8''$
(PA=52$^0$) and contours set at -0.5, 0.5, 1, 1.5, 2, 3, 4, 5, 6, 7, 8,
10, 15, 20, 25, 30, 35, 40, 50 mJy beam$^{-1}$ and a greyscale flux
range between
-2 and 6.6  mJy beam $^{-1}$; (m) The northern kink
of
the Snake with contour levels set at -0.5, 0.5, 1, 1.5, 2, 3, 4, 5, 6,
7,
8, 10, 15, 20, 25, 30, 35, 40, 50 mJy beam$^{-1}$ and a greyscale flux
range between
-1.1 and 10.30  mJy beam $^{-1}$; (n) Contours at 2,
4,
10, 30, 40 mJy beam$^{-1}$ and a greyscale flux range between
-1 and 3  mJy beam $^{-1}$; (o)
Contours of RF-C18 with levels -0.5,
0.5,
0.75, 1, 1.25, 1.5, 2, 2.5, and 3 mJy beam$^{-1}$ and a greyscale flux
range between
-3 and 15  mJy beam $^{-1}$; (p) Contours of RF-C6
and C7 with levels 1.5, 2, 2.5, 3, 3.5, 4, 5, 6, 7, 8, 10, 12, and 14
mJy
beam$^{-1}$ and a greyscale flux range between
-1 and 20  mJy beam $^{-1}$; (q) Contours of RF-C14
with levels -1, 0.5, 1, 1.5, 2, 2.5,
3, 4, 5, 6, 8, and 10 mJy beam$^{-1}$ and a greyscale flux range between
-1 and 12  mJy beam $^{-1}$;
(r) Contour levels of -1.5, 1.5,
2,
2.5, 3, 4, 5, and 6 mJy beam$^{-1}$ and a greyscale flux range between
-1 and 8  mJy beam $^{-1}$;
(s) Contour levels of -0.5, 0.5, 1
 and 1.5  mJy beam$^{-1}$ and a greyscale flux range between
-44 and 297.9 mJy beam $^{-1}$} \end{figure}

\begin{figure}
\figcaption{(a) 
Contours of emission from the Mouse G359.23-0.82 are set at -0.5, .5, 1,
1.5, 2, 3,
4, 5, 7, 9, 11, 13, 15, 20, 25, 30, 40, 50, 60, 70 and 90 mJy
beam$^{-1}$ and a greyscale flux range between
-2.7 and 100  mJy beam $^{-1}$;  (b) grayscale image of
G359.28-0.26 with contours -0.5,
.5, 1, 1.5, 2, 3, 4, 5, 6, 7. 8. 10, 15, 20, 25, 30, 35, 40 and 50 mJy
beam$^{-1}$ and a greyscale flux range between
-1.67 and 60  mJy beam $^{-1}$ (c) Contours of RF-C25
with levels 1, 1.5, 2, 4, 5, 10, 14,
20, and 26 mJy beam$^{-1}$ and a greyscale flux range between
--5 and 15  mJy beam $^{-1}$; (d)
Contours of RF-C21 and C19 with levels
0.5, 1, 1.5, 2, 3, 4, 6, 8, 10, and 12 mJy beam$^{-1}$
and a greyscale flux range between
-1.7 and 13  mJy beam $^{-1}$. Extended
emission from G359.16-0.04 is found at the terminus of the Snake (C19).
} \end{figure}

\begin{figure}
\figcaption{ 
(a) A mosaic image of Sgr E showing the distribution of HII regions in 
this 
nebula with a resolution of 11.3$\times7.9''$ (PA=54$^0$);
(b) a grayscale image of the central region of Sgr E with a resolution 
of 30$''\times30''$.
}\end{figure}

\begin{figure}
\figcaption{(a) Contours of 20cm emission from RF-E1 set 
at 0.5, 1, 2, 3, 5, 10, 15, 20, 25, 30, 35, 40, 45, 
50, 60, 70 and 90 mJy
 beam$^{-1}$ with a resolution of
 11.3$''\times7.9''$ (PA=54$^0$) and a greyscale flux range between
-1.7 and 10 mJy beam $^{-1}$; 
(b) a larger view of Sgr E and RF-E2 with contours similar to (a)
and a greyscale flux range between
-9 and 5  mJy beam $^{-1}$. 
}\end{figure}
\begin{figure}
\end{figure}

\begin{figure}
\figcaption{ (a) The vertical filaments (RF-S0) and the curved HII 
regions (i.e. the Arched filaments, the Sickle and the Pistol)  
comprise the Radio Arc at G0.2-0.0; the resolution is 12$''\times11''$.
The straight line at the bottom of the image is an artifact resulting
from combing two adjacent fileds. 
(b)  
The region to the south of the radio Arc showing 
filamentary structure in the southern lobe. The uv data $>$
0.4k$\lambda$ is selected. The resolution 
is 12.9$''\times9.2''$ (PA=73$^0$); (c) A second
field in the southern lobe 
with a resolution of 
10.9$''\times8.7''$ (PA=87$^0$).}
\end{figure}
\clearpage


\begin{figure}
\figcaption{ (a) A 20cm image of the southern extension of the radio Arc 
with 12$''\times9.4''$ (PA=64$^0$); (b) Contours at -2, 2, 3, 4, 5, 6,
7, 8, 9,
and 10 mJy beam$^{-1}$ and a greyscale flux range between
-3 and 10  mJy beam $^{-1}$. RF-S6 can be
seen clearly in Figure 16a; 
(c) Contours at -2, 2, 3, 4, 5, 6, 7, 8, 10, 12, 14, 16, 20,
25, and 30 mJy beam$^{-1}$ and a greyscale flux range between
-3 and 6  mJy beam $^{-1}$; (d)
Grayscale image with a 
resolution of 7.76$''\times6.3''$ (PA=66$^0$). The data includes 
DnC, CnB and BnA-array data; (e)
Similar to (d) but contours are shown at
2, 4, 6, 8,
10, 15, 20, 25, 30, 40, and 50 mJy
beam$^{-1}$ and a greyscale flux range between
-1.2 and 10  mJy beam $^{-1}$;  (f) Sources G0.38+0.02
and G0.42-0.32 can be noted
along
the extensions of RF-S2 (g) Contours at -3,
3, 4, 5, 6, 7, 8, and 10 mJy beam$^{-1}$
with the same resolution of (a) and a greyscale flux range between
-5 and 20  mJy beam $^{-1}$
(h) Contours at -2, 2, 4, 6,
10, 14, 18, 22, 30, 40, 50, 60, 80, and 100 mJy beam$^{-1}$ and a
greyscale flux range between
-5 and 20  mJy beam $^{-1}$; (i)
Contours at -2, 2, 3, 4, 5, 6, 7, 8, 10, 12, 14, 16, 20, 25, and 30 mJy
beam$^{-1}$ and a greyscale flux range between
-5 and 15  mJy beam $^{-1}$;  (j) Contours at -6, 6,
9, 12, 15, 18, 21, 24, 30, 36, 42,
48, 60, 75, and 90 mJy beam$^{-1}$ and a greyscale flux range
between
-3 and 20  mJy beam $^{-1}$;(k-l) Grayscale images with a
resolution of 10.7$''\times7.7''$ (PA=70$^0$); (17m) Contours at -4, -3, 2, 3, 4, 5, 10, 15, 20, 30, 40 and 50 mJy beam $^{-1}$ and a greyscale flux range between
-10 and 50  mJy beam $^{-1}$; (17n) Contours at 15, 20, 25, 30, 35, 40, 50, 60, 70, 80, and 100 mJy beam $^{-1}$ and a greyscale flux range
between
-17.8 and 50  mJy beam $^{-1}$; 
(17o) Greyscale image with a resolution of 3.5$''\times2.9''$ (PA=26$^{o}$) ; 
(17p) Contours at 2, 3, 4, 5, 6, and 8 mJy beam $^{-1}$ and a greyscale flux
range between
-2 and 6  mJy beam $^{-1}$; 
(17q) Greyscale image with a resolution of 2.1$''\times1.2''$ (PA=8$^{o}$) ; 
(17r) Contours at 0.5, 1, 2, 3, 4, 5, and 6 mJy beam $^{-1}$ and a greyscale flux
range between -0.5 and 3  mJy beam $^{-1}$. 
}
\end{figure}

\begin{figure}
\figcaption{(a) Contours at -1, 0.5, 1, 2, 3, 4, 5, 6, 8, 10, 12, 14, 16, 20, 
25, 30, 35, 40, 50, 60, and 70 mJy beam$^{-1}$ and a greyscale flux
range between
-3 and 10  mJy beam $^{-1}$;
(b) Contours at -5, 2.5, 5, 10, 15, 20, 25, 30, 40, 50, 60, 70, and 80 mJy
beam$^{-1}$ and a greyscale flux range between
-10 and 50  mJy beam $^{-1}$. 
(c) Contours at -1, 0.5, 1, 2, 3, 4, 5, 6, 8, 10, 12, 14, 16, 20, 25, 
30, 35, 40, 50, 60, 70, and 90 mJy beam$^{-1}$ and a greyscale flux
range between
-1 and 25  mJy beam $^{-1}$;  
(d) Contours at -1, 0.5, 1, 2, 3, 4, 5, 6, 8, 10, 
and 12 mJy beam$^{-1}$ and a greyscale flux range between
-0.3 and 8.9  mJy beam $^{-1}$. 
(e) Contours at -1, 0.5, 1, 2, 3, 4, 5, 6, 8, 10, 12, 14, 16, 
20, 25, 30, 35, 40, 50, 60, 70, and 90 mJy beam$^{-1}$ and a greyscale
flux range between
-1.6 and 10  mJy beam $^{-1}$;  
(f) Contours at -3, 3, 4, 5, 7, 9, 
11, 13, 15, 20, 25, 30, 40, and 50 mJy beam$^{-1}$ and a greyscale flux 
range between
-10 and 60  mJy beam $^{-1}$. 
The resolution for (a-e) is 10.6$''\times9.3''$ (PA=9$^0$)
whereas for (f) is 10.3$''\times8.5''$ (PA=-80$^0$)} 

\end{figure}

\begin{figure}
\figcaption{(a-d) Grayscale images with a resolution of 30$''\times30''$,
3.5$''\times3.2''$ (PA=48$^0$), 
3.76$''\times3.21''$ (PA=67$^0$), 30.2$''\times25.3''$ (PA=62$^0$), 
respectively}
\end{figure}

\begin{figure}
\figcaption{
(a) Contours at -22.5, -17.5, -12.5, 
-7.5, -2.5, 2.5, 7.5, 
12.5, 17.5, 22.5, 27.5, 37.5, 50, 
62.5, 75, 100, 125, 150, 175, 200, and 225 mJy beam$^{-1}$ and a
greyscale flux range between
-100 and 300  mJy beam $^{-1}$. 
(b) A grayscale image with a resolution of 
30$''\times15''$ (PA=0$^0$);
(c) Contours at 1, 2, 4, 6, 8, 10, 12, 14, 16, 18, 20, 24, and 28 mJy beam$^{-1}$
with a resolution of 
7.3$''\times6.7''$ (PA=19$^0$) and a greyscale flux range between
-6 and 20  mJy beam $^{-1}$;
(d) Contours at 2.5, 5, 10, 15, 20, 
25, 35, 45, 55, 65, 75, 85, 95, 105, 125, 150, 175, and 200 mJy
beam$^{-1}$ and a greyscale flux range between
-3 and 6  mJy beam $^{-1}$;
(e) Contours at 1, 2, 3, 4, 5, 7, 9, 11, 13, 
15, 17, 19, 21, 25, 30, 35, 40, 50, and 60 mJy beam$^{-1}$ with a resolution of 
3.8$''\times2.8''$ (PA=72$^0$) and a greyscale flux range between
-3 and 10  mJy beam $^{-1}$;
(f) Grayscale image with a resolution of   
3.8$''\times2.8''$ (PA=72$^0$);
(g) Contours at -0.5, 0.5, 1, 
1.5, 2, 2.5, 3, 4, 5, 6, 8, 10, 13, 16, 19, 22, 25, and 30 mJy beam$^{-1}$
with a resolution of 
7.3$''\times6.7''$ (PA=19$^0$) and a greyscale flux range between
-5 and 10  mJy beam $^{-1}$;
(h) Contours at 1.5, 2, 2.5, 3, 4, 5, 6, 8, 10, 13, 16, 19, 22, 25, and 30 mJy
 beam$^{-1}$ with a resolution of 
7.3$''\times6.7''$ (PA=19$^0$) and a greyscale flux range between
-2.1 and 5  mJy beam $^{-1}$;
(i) Grayscale image with a resolution of 6.9$''\times6.4''$ (PA=15$^0$) ; 
(j) Contours at 2, 3, 4, 5, 6, and 8 mJy beam $^{-1}$ and a greyscale flux range
between -1.5 and 3  mJy beam $^{-1}$; 
(k) Grayscale image with a resolution of 2.1$''\times1.2''$ (PA=8$^0$) 
}
\end{figure}


\begin{figure}
\figcaption{(a-e) Grayscale images of Sgr A with a resolution of
2.2$''\times1''$ (PA=9.7$^0$) based only on the A-array data, 
3.5$''\times2.9''$ (PA=26$^0$)
 using the uv range of between and 3 and 50k$\lambda$
of the A-array data, 
3.5$''\times2.9''$ (PA=26$^0$) 
using the combined  A, B, C and D-array data, 
10.7$''\times7.7''$ 
(PA=70$^0$) and 0.7$''\times7.7''$ (PA=70$^0$), respectively;
(f-g) Grayscale images of the region between Sgr A and Sgr C with 
a resolution of 11.9$''\times8.9''$ (PA=59$^0$)
}
\end{figure}

\begin{figure} 
\figcaption{ (a) Contours at -5, 5, 7, 9, 11, 15, 20, 25,
30, 40, 60, 80, 100, 150, 200, 300,
500, 700 and 900 mJy beam$^{-1}$ with a resolution of 
10.7$''\times7.7''$ (PA=70$^0$) and a greyscale flux range between
-19 and 100  mJy beam $^{-1}$; (b)
Contours at -4, 4, 5, 6, 7, 8, 10, 14, 18, 22, 30, 40,
50, 60, 70, 80, 100, 120, 140, 160, 180 and 200 mJy beam$^{-1}$
with a resolution of 
3.5$''\times2.9''$ (PA=26$^0$) and a greyscale flux range between
-3.9 and 45 mJy beam $^{-1}$; (c)
Contours at -2, 2, 2.5,
 3.5, 4, 5, 7, 9, and 11 mJy beam$^{-1}$ with a resolution of 
3.5$''\times2.9''$ (PA=26$^0$) and a greyscale flux range between
-0.95 and 7.2  mJy beam $^{-1}$; (d)
Contours at -4, 4, 5, 6,
7, 8, 10, 14, 18, 22,
30, 40, 50, 60, 70, 80, 100, 120, 140, 160, 180, and 200 mJy beam$^{-1}$
with a resolution of 
3.5$''\times2.9''$ (PA=26$^0$) and a greyscale flux range between
-1.2 and 18.6  mJy beam $^{-1}$; 
(e) Contours at -5, 5, 6.25, 7.5, 8.75, 10, 12.5, 17.5, 22.5, and 37.50 mJy
beam$^{-1}$ with a resolution of  
3.5$''\times2.9''$ (PA=26$^0$) and a greyscale flux range between
0.35  and 19.34  mJy beam $^{-1}$; (f)
Contours at 0.5, 1.5, 2.5, 3.5, 5, 6.5, 8, 10, and 12
mJy beam$^{-1}$ with  a
resolution of 2.2$''\times1''$ (PA=9.7$^0$) and a greyscale flux range
between
-5.6 and 20  mJy beam $^{-1}$; (g)
Contours at 0.25, 0.75, 1.25, 1.75, 2.5, 3.25, 4, 5, 6, 7.5 and 10 mJy beam$^{-1}$ with  a
resolution of 2.2$''\times1''$ (PA=9.7$^0$) and a greyscale flux range
between
-3 and 14  mJy beam $^{-1}$; (h) Contours at 2, 3, 4, 7, 10, 15, 20, and 25 mJy beam$^{-1}$ and a greyscale flux range between -3 and 5  mJy beam $^{-1}$.}
\end{figure}

\begin{figure}
\figcaption{
(a) Contours at 0.5, 1, 2, 4, 6, 8, 10, 12, 14, 16, 19, 20, 30, 40 and 50
 mJy beam$^{-1}$ and a greyscale flux range between
-5 and 10  mJy beam $^{-1}$;
(b) Contours at 2, 3, 4, 5, 6, and 7 mJy beam$^{-1}$ and a greyscale
flux range between
-2 and 15  mJy beam $^{-1}$;
(c) Contours at -2, 2, 2.5, 3, 3.5, 4, 5, 7, 9, and 
11 mJy beam$^{-1}$; the resolution 
of all images is 2.2$''\times1''$ (PA=9.7$^0$) and a greyscale flux
range between
-1.4 and 15.7  mJy beam $^{-1}$}
\end{figure}

\begin{figure}
\figcaption{(a) grayscale image of Sgr B with a resolution of
2.52$''\times1.7''$ (PA=21$^0$) based on combining BnA, C and D-array data using the masking
technique; The same region as (a) except that only BnA-array data is used having the same  
resolution; (c) only C and D array data used with a 
resolution of 18$\times18''$; (d) similar to (a) }
\end{figure}

\begin{figure}
\figcaption{
(a) Contours at 3.5, 7, 10.5, 14, 21, 28, 35, 
49, 63, 77, 105, 
140, 175, 210, 280 and 350 mJy beam$^{-1}$ with the same resolution as
Figure 24c and a greyscale flux range between
-16 and 500  mJy beam $^{-1}$;
(b-c) A Grayscale image and contours  at 1, 3, 5, 7, 9, 11, 15, 20, 25, 30, 40,
50, 70, 90 and
110 mJy beam$^{-1}$; (d-e) A gryscale image and its  contours at 1.5, 2.5, 
3.5, 4.5, 5.5, 7.5, 10, 12.5, 15, 20, 25,
35, 45 and 55 mJy beam$^{-1}$;
(f-g) A frayscale image and its contours 
at 1.5, 2.5, 3.5, 4.5, 5.5, 7.5, 10, 12.5, 15, 20,
25, 35, 45 and 55 mJy beam$^{-1}$. The resolution of (b)  to (e) 
is the same as  Figure 24b. }
\end{figure}

\begin{figure}
\figcaption{A high resolution mosiac image of eight 20cm fields
with a resolution of 10$''\times10''$}
\end{figure}

\begin{figure}
\figcaption{
(a) Contours of 
SNR G0.9+0.1 with levels at 2.5, 5, 7.5, 10, 
15, 20, 25, 35, 45, 55, 75, 100, 
125, and 150 mJy beam$^{-1}$. 
The  beam size is  11.3$''\times8.6''$ (PA=-74$^0$) and a greyscale
flux range between
-5 and 25  mJy beam $^{-1}$;  
(b) Contours of SNR G1.4-0.1 with levels at 0.75, 1.5, 2.25, 3, 4.5, 6, 
7.5, 10.5, 13.5, 16.5, 22.5, 30, 37.5, and 45 mJy beam$^{-1}$
The  beam size is  12.2$''\times8.2''$ (PA=65$^0$) and a greyscale flux 
range between
-1 and 3  mJy beam $^{-1}$;
(c) Contours of the northern shell of SNR G3.7-0.2 with levels at 0.5, 1, 
1.5, 2, and 3 mJy beam$^{-1}$ and a greyscale flux range between
-0.5 and 3  mJy beam $^{-1}$. 
The  beam size is  11.4$''\times-8.4''$ (PA=-84$^0$); 
(d) Contours of SNR G1.9+0.3 with levels 
at 1, 2, 3, 4, 6, 8, 10, 12, 14, 18, 22, 30, 
40, 50  and 60 mJy beam$^{-1}$
The  beam size is  12$''\times8.3''$ (PA=-62$^0$) and a greyscale flux
range between
-2 and 50  mJy beam $^{-1}$;
(e) Contours of Sgr D SNR with levels at
9, 12, 15, 18, 21, 24, 27, 30, 35, 40, 45, 50, 60, 70)$\times$2 mJy
beam$^{-1}$ with a resolution of 30$\times30''$ and a greyscale flux
range between
4.6 and 100  mJy beam $^{-1}$. The single-dish Bonn
data has been folded  into  this image  
(f) Contours of SNR candidate G3.66-0.12 with levels at 
1, 2, 3, 4, 6, 8, 10, 14, 18, 22, 30, 40, 50, and 60 mJy beam$^{-1}$.
The  beam size is  12.7$''\times8.5''$ (PA=--78$^0$) and a greyscale
flux range between
-1.7 and 10 mJy beam $^{-1}$. 
}
\end{figure}

\begin{figure}
\figcaption{(a) Contours at 1, 2, 3, 4, 5, 6, 7, 8, 10, 14, 18, 
22, 30, 40, 50, 60 mJy
beam$^{-1}$. 
The central structure G1.32+0.09 with an extent of  
2$'\times4'$ and a total integrated 
flux  of 1.1 Jy. 
The  beam size is  12.2$''\times8.2''$ (PA=65$^0$) and a greyscale flux 
range between
-4 and 20  mJy beam $^{-1}$.
(b) Contours at 0.5, 1, 1.5, 2, 3, 4, 5, 7, 
9, 11, 15, 20, 25, and 30 mJy beam$^{-1}$.
The  beam size is  11.8$''\times8.8''$ (PA=82$^0$) and a greyscale flux 
range between
-1 and 20  mJy beam $^{-1}$;
 (c) Contours at 0.5, 1, 1.5,
2, 3, 4, 5, 7, 9, 11, 15, 20, 25, and 30 mJy beam$^{-1}$.
The  beam size is  11.8$''\times8.8''$ (PA=82$^0$) and a greyscale flux 
range between
-1 and 20  mJy beam $^{-1}$;
(d) Contours of HII region G4.57-0.12 with levels at 1, 1.5, 2, 3, 4, 5,
7, 9, 11, 15, 20, 25, and 30 mJy beam$^{-1}$. 
The  beam size is  11$''\times8.2''$ (PA=-88$^0$) and a greyscale flux
range between
-3 and 10  mJy beam $^{-1}$;
 (e) Contours of HII region
G4.42+0.12 with levels
at 0.5, 1, 1.5, 2, 3, 4, 5, 7, 9, 11, 15, 20, 25, and 30 mJy beam$^{-1}$. 
The  beam size is  11$''\times8.2''$ (PA=-88$^0$) and a greyscale flux
range between
-1 and 20  mJy beam $^{-1}$;
(f) Contours of SNR candidate G4.27+0.04 with levels 0.5, 1, 1.5, 2, 3, 4,
5, 7, 9, 11, and 15 mJy beam$^{-1}$.
The  beam size is  11$''\times8.2''$ (PA=-88$^0$) and a greyscale flux
range between
-1.5 and 4  mJy beam $^{-1}$;
 (g) Contour levels at
0.5, 1, 1.5, 2, 3, 4, 5, 7, 9, 11, 15, 20, 25, and 30 mJy beam$^{-1}$.
The  beam size is  11$''\times8.2''$ (PA=-88$^0$) and a greyscale flux
range between
-1.5 and 4  mJy beam $^{-1}$;
(h)
Contour levels at 0.5, 1, 1.5, 2, 3, 4, 5, 7, 9, 11, 15, 20, 25, and 30
mJy beam$^{-1}$.
The  beam size is  12.7$''\times8.5''$ (PA=-77$^0$) and a greyscale
flux range between
-1.7 and 10  mJy beam $^{-1}$;
 (i) Contour levels at 0.5, 1, 1.5, 2, 3, 4, 5, 7, 9, 11,
15, 20, 25, and 30 mJy beam$^{-1}$. 
The  beam size is  12.7$''\times8.5''$ (PA=-77$^0$) and a greyscale
flux range between
-1.5 and 2  mJy beam $^{-1}$;
 (j) Contours of an HII 
complex
with levels at 1, 2, 3, 4, 6, 8, 10, 14, 18, 22, 30, 40,
50, and 60 mJy beam$^{-1}$.
The  beam size is  12.7$''\times8.5''$ (PA=-77$^0$) and a greyscale
flux range between
-1.7 and 10  mJy beam $^{-1}$;
(k) Contour levels at 0.5, 1, 1.5, 2, 3, 4, 5,
7, 9, 11, 15, 20, 25, and 30 mJy beam$^{-1}$.
The  beam size is  11.1$''\times8.4''$ (PA=-75$^0$) and a greyscale
flux range between
-0.5 and 2  mJy beam $^{-1}$;
(l) Contour levels at 0.5,
1, 1.5, 2, 3, 4, 5, 7, 9, 11, 15, 20, 25, and 30 mJy beam$^{-1}$. 
The  beam size is  11.1$''\times8.4''$ (PA=-75$^0$) and a greyscale
flux range between
-1 and 5 mJy beam $^{-1}$;
(m)
Contour levels at 0.5, 1, 1.5, 2, 3, 4, 5, 7, 9, 11, 15, 20, 25, and 30
mJy beam$^{-1}$. 
The  beam size is  11.8$''\times8.8''$ (PA=82$^0$) and a greyscale flux 
range between
-1 and 5  mJy beam $^{-1}$;
 (n) Contours of HII region
G2.30+0.24 with levels at 1, 2, 3, 4, 6, 8, 10, 14, 18, 22, 30, 40, 50,
and 60 mJy beam$^{-1}$. 
The  beam size is  11.8$''\times8.8''$ (PA=82$^0$) and a greyscale flux 
range between
-1 and 20  mJy beam $^{-1}$} 
\end{figure} 
\clearpage

\begin{figure} 
\figcaption{ The
schemactic diagram of all the identified radio filaments labeled 
in this survey. The position of Sgr A$^*$ is presented by a star. 
The light background circles show the angular size of the 
surveyed region. 
There is also a NRF G358.85+0.47 extending for 7$'$ in its
length and runs along the Galactic plane (Lang et al. 1999). This
NRF was not covered by our survey but  is shown in this diagram. 
}
\end{figure} 
\clearpage

\end{document}